\documentclass[apj,twocolumn,twocolappendix,dvipsnames]{aastex63}

\usepackage{epstopdf}
\usepackage{pbox}
\usepackage{gensymb}
\usepackage{amsmath}
\usepackage{amssymb}

\DeclareSymbolFont{tensors}{OT1}{cmss}{bx}{it}
\DeclareSymbolFontAlphabet{\mathtens}{tensors}

\renewcommand{\vec}[1]{\mathbf{#1}}
\newcommand{\tens}[1]{\mathsf{#1}}
\newcommand{\pd}[2]{\frac{\partial #1}{\partial #2}}
\newcommand{\DS}{\displaystyle}

\begin{document}
\title{
Jets from accretion disk dynamos: consistent quenching modes for dynamo and resistivity
}

\author[0000-0003-1454-6226]{Giancarlo Mattia}
\altaffiliation{Member of the International Max Planck Research School for Astronomy \& Cosmic Physics at the University of Heidelberg}
\author[0000-0002-3528-7625]{Christian Fendt}
\affiliation{Max Planck Institute for Astronomy, Heidelberg, Germany}


\email{mattia@mpia.de, fendt@mpia.de}



\begin{abstract}
Astrophysical jets are launched from strongly magnetized systems that host an accretion disk surrounding a central object. 
The origin of the magnetic field, which is a key component of the launching process, is still an open question. 
Here we address the question of how the magnetic field required for jet launching is generated and maintained by a dynamo process. 
By carrying out non-ideal MHD simulations (PLUTO code), we investigate how the feedback of the generated magnetic field on the mean-field 
dynamo affects the disk and jet properties. 
We find that a stronger quenching of the dynamo leads to a saturation of the magnetic field at a lower disk magnetization. 
Nevertheless, we find that, while applying different dynamo feedback models, the overall jet properties 
remain unaffected.
We then investigate a feedback model which encompasses a quenching of the magnetic diffusivity. 
Our modeling considers a more consistent approach for mean-field dynamo modeling simulations, as the magnetic quenching of 
turbulence should be considered for both, a turbulent dynamo and turbulent magnetic diffusivity. 
We find that, after the magnetic field is saturated, the Blandford-Payne mechanism can work efficiently, leading to more 
collimated jets, that move, however, with slower speed. 
We find strong intermittent periods of flaring and knot ejection for low Coriolis numbers. 
In particular, flux ropes are built up and advected towards the inner disk thereby cutting off of the inner 
disk wind, leading to magnetic field reversals, reconnection and the emergence of intermittent flares.
\end{abstract}

\section{Introduction}
%
Astrophysical jets, that are highly-collimated beams of high-velocity material, are ejected from a variety of sources, such as
young stellar objects (YSOs), micro-quasars (MQ), or active galactic nuclei (AGNs).
They thus span a wide range in terms of extension, timescale and energy output, 
but there is common understanding that jet sources also share common properties that are responsible for jet launching.
That is the presence of a central object, which is surrounded by an accretion disk 
(see e.g. \citealt{2014prpl.conf..451F, 2015SSRv..191..441H, 2019FrASS...6...54P}).
Another fundamental prerequisite for the formation of a jet is the existence of a strong, large-scale magnetic field.
All launching processes that have been proposed through the decades (see e.g. \citealt{1977MNRAS.179..433B,1982MNRAS.199..883B,1996MNRAS.279..389L}), 
rely on the presence of a strong magnetic field with a favorable topology. 
This holds for both the non-relativistic and the relativistic limit.

The jet launching process can be studied by magnetohydrodynamic (MHD) simulations
(see e.g. \citealt{1985PASJ...37..515U,2002ApJ...581..988C,2006ApJ...651..272F, 2007A&A...469..811Z,2014ApJ...793...31S}),
mostly assuming a large-scale, strong, initial magnetic field.
While this approach allows to understand a number of properties of the jet that is formed, 
it does not address the question about the formation of such a magnetic field.
Several studies have pointed out a strong correlation between the disk magnetic field and the jet properties \citep{2009MNRAS.400..820T,2010A&A...512A..82M,2016ApJ...825...14S}. 
Therefore, the understanding of the origin of the disk magnetic field represents a key step for understanding the formation 
of astrophysical jets.

For different central objects, several options for the magnetic field origin may be considered.
For instance, in case of a {\em stellar} accreting object, the magnetic field may be provided by the star itself.
This scenario is not feasible for AGN, since the black hole is not able to generate a magnetic field.
Another scenario is that the jet-launching magnetic field is just advected through the accretion process
from the ambient medium.
Still, for protostars, a strong evidence for the advection of magnetic flux has not been found yet \citep{2019FrASS...6...54P}.
Therefore, a particularly interesting scenario is that the magnetic field can be generated and amplified by dynamo process that is 
working in the accretion disk.
Disks are turbulent and in rotation, providing the two main ingredients for a dynamo (see below).

In this paper, we will indeed focus on the disk dynamo mechanism, because it presents a very general mechanism 
that can be responsible for generating a large magnetic flux in jet-launching accretion disks in both the 
protostellar and the AGN context.
In particular, if we consider AGN, the disk magnetic field can also provide the magnetic flux that is needed to
launch jets via the Blandford-Znajek mechanism.

The physics of cosmic dynamo action has been extensively studied in the last decades, in particular by numerical simulations
(see e.g. \citealt{2005PhR...417....1B,2019JPlPh..85d2001R} 
for reviews on dynamo theory and their astrophysical applications).
Astrophysical accretion disk dynamos are believed to have a turbulent origin, most probably caused by the magneto-rotational instability, MRI \citep{1991ApJ...376..214B,1991ApJ...376..223H}.
The dissipation of the magnetic field has also been modeled in the same fashion (see, e.g., \citealt{2002ApJ...581..988C}).

Accretion disk dynamos have been suggested already decades ago
\citep{1981MNRAS.195..881P,1981MNRAS.195..897P,1995ApJ...446..741B} in order to describe the strong magnetic field in turbulent accretion disks around black holes. 
A comprehensive modeling of jet launching by a dynamo-generated magnetic field is a challenging task because of the 
different physical mechanisms that operate over a very wide range of spatial and temporal scales.
The spatial scale stretches from the scale on which turbulence occurs, to the scale on which the jet is launched.
On the small scales, resolution studies of the MRI have been performed by a number of groups
(see e.g. \citealt{2009ApJ...697.1901G,2010ApJ...713...52D,2013ApJ...772..102H,2017ApJ...840....6R}).
Modeling the disk and jet regions at the turbulent scales would be computationally hardly feasible, requiring
unrealistically high computational resources.

For all these reasons, mainly two different types of dynamo action have been investigated.
On one hand, so-called direct simulations show the direct amplification of the magnetic field by the turbulent motion
(see, e.g., 
\citealt{2010MNRAS.405...41G,2013ApJ...767...30B,2015ApJ...810...59G,2018MNRAS.474.2212R,2018ApJ...861...24H,2020MNRAS.494.4854D}).
On the other hand, there is the so-called mean-field dynamo approach 
(see, e.g., \citealt{1980opp..bookR....K,1995A&A...298..934R,1999MNRAS.310.1175C,2000A&A...353..813R, 2001A&A...370..635B,2006A&A...446.1027C}) 
that relies on (semi-)analytical solutions of the turbulent motion, 
allowing us to perform the large-scale numerical simulations required to consider the scales on which accretion disks and jets evolve.

In the present work we follow the mean-field approach.  
Whether a large-scale magnetic field can be generated by a mean-field dynamo has been numerically investigated by \citet{2003A&A...398..825V,2014ApJ...796...29S,2018ApJ...855..130F,2020ApJ...900...59M,2020ApJ...900...60M}.
These results have demonstrated that a dynamo-amplified, strong magnetic field can evolve and is able to launch a jet or a disk outflows
in general.
In particular, it has been shown by \citet{2004A&A...420...17V, 2018MNRAS.477..127D} how an accretion disk dynamo-generated 
magnetic field leads to an outflow also if combined with a central stellar magnetic field.
Recently, the mean-field dynamo approach has been also extended to general relativistic MHD simulations in torii
\citep{BdZ2013, Bugli2014, 2020MNRAS.491.2346T,2021Univ....7..259T,2022Fluid...7...87D} or in thin accretion disks \citep{2021ApJ...911...85V}.

In this paper we want to further bridge the gap between analytical models and the time-dependent simulations.
We focus on the non-linear evolution of the dynamo phenomena,
i.e. how a strong disk magnetic field that has been dynamo-generated affects itself 
the dynamo action and the diffusivity within the accretion disk, which have both a turbulent origin.

Despite several options to parameterize a mean-field dynamo and turbulent diffusivity being developed in
the literature, so far, based on analytical modeling of the disk turbulence
(see e.g. \citealt{1993A&A...269..581R,1994AN....315..157K,1994GApFD..78..247R}), 
a consistent numerical treatment of the feedback has never been adopted in the context of the jets launching
from a resistive and dynamo-active accretion disk.
We thus compare different feedback strategies adopted in the context of accretion disk dynamos and we investigate 
how the different feedback models affect the jet launching process.
Finally we investigate the effect of a - we think more consistent - diffusivity model which includes a quenching 
mode for strong magnetization.
The latter is motivated by the fact that both the alpha dynamo and the magnetic diffusivity applied are believed to
have the same source, namely the disk turbulence. 
Thus, a quenching of the turbulent dynamo by a strong field should also consider the quenching of turbulent dissipation
at the same time.
This has not yet been considered by thin-disk dynamo simulations launching jets.

The paper is organized as follows.
In Section \ref{Sec:model} we describe our numerical setup. 
In Section \ref{Sec::dyn_quench} we investigate the effects of different feedback dynamo models and their role in the jet launching process.
In Section \ref{Sec::eta_quench} we introduce a consistent turbulent model of both mean-field dynamo and magnetic diffusivity for thin accretion disks and we apply it to the large scale disk-jet simulation.
We summarize our paper in Section \ref{Sec:conclusions}.

\section{Model approach}
\label{Sec:model}

\subsection{MHD equations}
%
We solve the time-dependent, resistive MHD equations with the PLUTO code\footnote{http://plutocode.ph.unito.it/} \citep{2007ApJS..170..228M} 
version 4.3 in spherical coordinates $(R,\theta,\phi)$ assuming axisymmetry.
We refer to $(r,z)$ as cylindrical coordinates.
The code solves the integral form of the MHD conservation laws through a high order finite volume method.
To the standard PLUTO MHD equations we have added a mean-field dynamo term.

The set of primitive variables consist of, respectively, the plasma density $\rho$, 
the flow velocity ${\vec v}$,
the gas pressure $p$ and the magnetic field ${\vec B}$.

The set of conservation laws consists of the conservation of mass,
\begin{equation}
\pd{\rho}{t} + \nabla\cdot(\rho\vec{v}) = 0,
\end{equation}
momentum,
\begin{equation}
    \pd{\rho\vec{v}}{t} + \nabla\cdot\left[\rho\vec{v}\vec{v}
    + \left(p + \frac{\vec{B}\cdot\vec{B}}{2}\right)\tens{I} 
    - \vec{B}\vec{B}\right]
    + \rho\nabla\Phi_{\rm g} = 0,
\end{equation}
total energy,
\begin{equation}
    \pd{e}{t} + \nabla\cdot\left[\left(\DS\frac{\rho v^2}{2} + \DS\frac{\Gamma}{\Gamma-1}p + \rho\Phi_{\rm g}\right){\vec v} + {\vec E}\times{\vec B}\right] 
              = \Lambda_{\rm{cool}},
\end{equation}
and magnetic field,
\begin{equation}
    \pd{\vec{B}}{t} + \nabla\times{\vec E} = 0.
\end{equation}
The total energy density is computed, from the primitive variables, as
\begin{equation}
    e = \frac{p}{\Gamma - 1}
      + \DS\frac{\rho v^2}{2}
      + \DS\frac{B^2}{2}
      +  \rho\Phi_{\rm g},
\end{equation}
with polytropic index $\Gamma = 5/3$,
while the electric field is defined by (see \citealt{1980opp..bookR....K})
\begin{equation}
    {\vec E} = -{\vec v}\times{\vec B} + \mathtens{\eta}\cdot(\nabla\times{\vec B}) - \mathtens{\alpha}\cdot{\vec B}
\end{equation}
where the first is the ideal term, the second is the diffusive term and the third is the dynamo term.
The gravitational potential $\Phi_{\rm g}$ is provided by a central object mass M, i.e. $\Phi_{\rm g} = -GM/R$.

Previous studies (see e.g. \citealt{2000A&A...353.1115C,2000A&A...361.1178C,2007A&A...469..811Z,2013MNRAS.428.3151T}) have shown that the non-ideal heating and cooling processes may play a relevant role in the context of jet formation. 
Here, for the sake of simplicity, the cooling term $\Lambda_{\rm{cool}}$ is set to be equal to the non-ideal (i.e. diffusive and dynamo) contribution of the electric field to the energy equation, as in \citet{2012ApJ...757...65S,2014ApJ...793...31S}:

\begin{equation}
    \Lambda_c = \nabla\cdot\{[\mathtens{\eta}\cdot(\nabla\times {\vec B}) - \mathtens{\alpha}\cdot{\vec B}]\times{\vec B}\}
\end{equation}

We choose to neglect the non-ideal heating process in accordance with previous studies of jet launching from resistive disks \citep{2002ApJ...581..988C}.

The tensors $\mathtens{\alpha}$ and $\mathtens{\eta}$ describe the effects of the mean-field dynamo and the magnetic diffusivity on the magnetic field evolution (see Sections \ref{sec::eqdynamo} and \ref{sec::eqdiffusivity}).

\subsection{Numerical setup}
%
Since the PLUTO code solves the MHD equation in their adimensional form, intrinsic physical scales are not involved.
We apply code units, for which every scale (time, length, energy, etc.) is normalized to its initial 
value at the innermost disk radius $R_{\rm in}$ at the midplane.
Therefore, velocities and are normalized to $v_{\rm k,in}$, the Keplerian speed at $R_{\rm in}$.
As a consequence, the time is normalized to $t_{\rm in} = R_{\rm in}/v_{\rm k,in}$.
Thus, $2\pi t_{\rm in}$ corresponds to one disk revolution at $R_{\rm in}$.
For the sake of simplicity, all times that are shown without a normalization constant, are measured in units
of $t_{\rm in}$.

The computational domain ranges in the radial direction between $R\in[1,100]R_{\rm in}$, and in the angular direction, 
between $\theta\in[10^{-8},\pi/2-10^{-8}]\simeq[0,\pi/2]$ (where $\theta = 0$ corresponds to the jet axis and $\theta = \pi/2$ corresponds to the disk midplane).
In order to be able to cover a large extension of the numerical grid, we apply stretched cells in the radial direction 
($\Delta R = R\Delta\theta$), while the spacing in the angular direction is equidistant.
We adopt a resolution of $[N_R\times N_\theta] = [512\times256]$ through the whole paper. 
This gives a resolution throughout the paper of 32 cells per geometrical disk height $H$.
We found that a lower resolution may lead to numerical issues, especially in case of strong interplay between the magnetic 
field and the dynamo in the innermost accretion disk regions.

The spatial reconstruction is performed employing the piecewise parabolic method for spherical coordinates 
\citep{2014JCoPh.270..784M}, while the time integration is achieved through a third-order Runge-Kutta scheme.
The flux is computed through a Harten-Lax-van Leer Contact (HLLC) Riemann solver \citep{1994ShWav...4...25T,2005JCoPh.203..344L}
\footnote{In case a simulation encounters convergence issues, we switch from the HLLC to the HLL Riemann solver 
\citep{1983Siam...25...25H,2009book.123..123}, in order to increase the stability of the numerical scheme}.

The solenoidality of the magnetic field is preserved through an upwind constrained transport method \citep{2004JCoPh.195...17L}.
In particular, as in \citet{2020ApJ...900...59M,2020ApJ...900...60M}, we employ a combination of the UCT-HLL for the ideal and 
resistive components, while the dynamo is added through an arithmetic average
(see \citealt{2021JCoPh.42409748M} for a more detailed study on the upwind constrained transport schemes).
In order to preserve stability, we adopt a Courant-Friedrichs-Levy time stepping with $CFL = 0.4< 1/\sqrt{N_{\rm dim}}$.

\subsection{Initial conditions}
\label{sec::init}
%
The initial conditions are identical to those applied in \citet{2020ApJ...900...59M,2020ApJ...900...60M}.
For the sake of clarity, we summarize the fundamental steps.
The simulation starts with a very weakly magnetized disk.
The initial magnetization, defined as the ratio between the thermal and the magnetic pressure both measured at the disk midplane 
is $\mu_{\rm in} = B_{\rm in}^2/2p_{\rm in} = 10^{-5}$. 

Therefore, we recover the initial disk structure by assuming equilibrium between thermal pressure gradients, gravity 
and centrifugal forces \citep{2014ApJ...796...29S, 2018ApJ...855..130F},
\begin{equation}
\label{eq::init_disk}
    \nabla p + \rho\nabla\Phi_{\rm g} - \DS\frac{\rho v_\phi^2}{R}({\vec e}_R\sin\theta + {\vec e}_\theta\cos\theta) = 0.
\end{equation}
This equation can be solved by assuming that all the hydrodynamical variables $X$ scale as power laws of the radius $R$, i.e. $X = X_0R^{\beta_X}F_X(\theta)$, 
where $X_0$ is the corresponding variable evaluated at the inner disk radius, while $F_X$ is the angular dependence.
We assume a polytropic gas, $p\propto\rho^\Gamma$, and that all characteristic velocities 
scale as the Keplerian velocity, $\beta_{v_\phi} = -1/2$.
Under these assumptions, the radial dependence of the hydrodynamical variables is determined by 
$\beta_\rho = -3/2$ and $\beta_p = -5/2$ \citep{1982MNRAS.199..883B}.

In order to recover the explicit value of $F_X(\theta)$ we need to set two parameters, both evaluated at the inner disk radius: 
the disk density $\rho_0$ and the thermal disk height, defined as the ratio between the isothermal sound speed and the Keplerian speed
$\epsilon = c_s/v_\phi|_{\theta = \pi/2}$.
We set $\rho_0 = 1$ and $\epsilon = 0.1$.

Once the density and pressure distributions are recovered, we use the combination of the radial and angular component of Eq.~\ref{eq::init_disk} 
in order to compute the angular dependence of the toroidal velocity 
(as in \citealt{2009A&A...508.1117Z,2013A&A...550A..99Z} but neglecting the viscous terms).

Above the disk we define an hydrostatic corona, 
\begin{equation}
\label{eq::init_corona}
\begin{array}{lcl}
    \rho_{\rm c} & = &
    \rho_{\rm{c,in}} R^{1/(1-\Gamma)}, \\ \noalign{\medskip}
     p_{\rm c} & = & p_{\rm{c,in}}R^{\Gamma/(1-\Gamma)}
     =\DS\frac{\Gamma-1}{\Gamma}\rho_{\rm{c,in}}R^{\Gamma/(1-\Gamma)},
     \end{array}
\end{equation}
with $\rho_{\rm{c,in}}=10^{-3}\rho_0$.
At the transition between accretion disk and corona the disk pressure equals the coronal pressure, involving a jump in density.

As pointed by \citet{2014ApJ...793...31S}, we define as geometrical disk height the region where density and toroidal velocity decrease significantly. 
Since through all the paper we assume $\epsilon = 0.1$, we adopt a linear approximation $H = 2r\epsilon$ which is able to reproduce with great accuracy the relation between the thermal and the geometrical disk height.
The study of the influence that the disk height has on the dynamo-generated magnetic field will be the subject of our future works.

All the simulations are initialized with a weak purely radial magnetic field defined by the vector potential \citep{2014ApJ...796...29S}
\begin{equation}
    \vec{B}=\nabla\times A\vec{e}_\phi=
    \nabla \times \left[ \frac{B_{\rm{p,in}}}{r}  \exp \left( -8 \left(z/H\right)^2 \right) \right]\vec{e}_\phi.
\end{equation}
The strength of the initial poloidal magnetic field is determined by $B_{\rm p,in} = \sqrt{2p_0\mu_{\rm in}}$, 
where $\mu_{\rm in} = 10^{-5}$ is the initial magnetization along the disk midplane.
By construction, the initial toroidal magnetic field $B_\phi$ vanishes.

\subsection{Boundary conditions}
%
The physical evolution is heavily determined by the choice of boundary conditions.
A {"}wrong{"} choice of boundary conditions may easily lead to a non consistent or unphysical scenarios
(see e.g. a recent review on these issues by \citealt{2021A&A...652A..38B}).

The boundary conditions we adopt here are a slightly revised version of the approach in \citet{2014ApJ...796...29S,2018ApJ...855..130F}, 
and are reported in Table \ref{tab::boundaries} for convenience.
Standard symmetry conditions are applied along the disk mid-plane and along the rotational axis.

Equatorial symmetry (i.e., in this case, symmetry for the component $B_\theta$ and anti-symmetry for the components $B_R$ and $B_\phi$) conditions does have certain consequences.
Concerning the magnetic field it imposes that the toroidal magnetic field does vanish along the equatorial plane.
Thus the poloidal field intersects the mid-plane perpendicularly.
While with our approach we constrain the disk mid-plane to the equatorial plane, we expect similar
effects also from a bipolar approach, unless a large-scale bipolar asymmetry will build up.
In \citet{2018ApJ...855..130F} the bipolar setup lead to some disk warping, however, on the large
scales the field remained symmetric. 
This implies that across the disk mid-plane (in that case different from the equatorial plane of the grid)
the toroidal field changes sign, leading to a vanishing toroidal field over there - similar to our present 
setup. 
For the choice of the sign of $\alpha$ (in both papers) a large scale flux emerged.

The situation may be different for mean-field dynamo of opposite sign.
This may be relevant in particular for bipolar simulations considering both hemispheres.
\citet{2000A&A...353..813R,2001A&A...370..635B} have shown that a change of sign for the $\alpha$ 
(applied to both hemispheres leads to a quadrupolar large scale magnetic field geometry.
These results, obtained by a semi-analytical treatment has been confirmed by numerical simulations
of a scalar mean-field dynamo, finding either dipolar (negative $\alpha$, upper hemisphere) or
quadrupolar fields (positive $\alpha$, upper hemisphere) \citep{2018ApJ...855..130F}. 
This effect can, obviously, not be investigated if only one hemisphere is considered.
However, such an investigation would be beyond the aim of the present paper, which emphasis is to apply
more consistent dynamo and quenching models.
A different polarity of the dynamo would also be in contrast with the analytical model of \citet{1995A&A...298..934R}, 
and we thus choose to not investigate further the polarity of $\alpha$.

Other reasons to constrain ourselves to one hemisphere is just simplicity.
We intend to investigate the pure dynamo effect that should not be affected by the 
dynamics of the disk mid-plane.

In fact, our prescription for the $\alpha$ and $\eta$ {\em distribution} is fixed in space
(and centered on the mid-plane)\footnote{Simulations with a local quenching - i.e. a quenching from the local physics
in the disk would avoid this problem, however, would also neglect the idea that the disk $\alpha$-turbulence is 
generated by (MRI) instabilities in the center of the disk and than transported upwards considering helicity.}.
To allow for disk warping would require to determine the {\em physical} disk mid-plane for each time step 
and update for the $\alpha$ and $\eta$ distribution consequently.
This has {\em not} been done previously and it is beyond the scope of the paper.

Both the inner and outer radial boundaries are separated into a disk $(\theta > \pi/2 - 3\epsilon)$ and a coronal 
$(\theta < \pi/2 - 3\epsilon)$ region.
The extent of the inner disk boundary is somewhat broader than the initial disk height.
The reason is that the disk, especially in case of strong accretion, may slightly inflate,
but all material delivered by disk accretion must be able to be absorbed by the boundary.

Along both the disk and the coronal inner boundary we prescribe $v_\theta = 0$.
For the disk part, density, pressure and toroidal velocity are extrapolated from the domain 
quantities by a power law.
The radial velocity is also computed by extrapolating a power law, 
however we allow only for negative radial velocity.

The density and pressure for the inner coronal boundary condition are set to their initial value, but
multiplied by a factor $\rho_{\rm c,in}$ (e.g. $p_{\rm c,in}$ for the pressure, see Eq. \ref{eq::init_corona}), and describing a fixed radial inflow of $v_R = 0.1$.
Together, this allows for a more stable evolution between the disk and coronal boundary
(compared to \citealt{2014ApJ...796...29S,2020ApJ...900...59M,2020ApJ...900...60M})
as strong density gradients may appear at the interface.

\begin{table*}
\caption{Shown is the choice of how the variables in the outermost domain cell
are extrapolated across the respective ghost cells of the boundary. {\em Outflow} denotes standard, zero-gradient 
PLUTO outflow boundary conditions. Alternatively, divergence-free conditions are involved, a prescribed field inclination,
or flux conservation.
\label{tab::boundaries} }
 \centering
\begin{tabular}{ccccccccc}
 \hline
  & $\rho$ & $p$ & $v_R$ & $v_\theta$ & $v_\phi$ & $B_R$ & $B_\theta$ & $B_\phi$ \\
 \hline 
 Inner disk   & $\propto R^{-3/2}$ & $\propto R^{-5/2}$ & $\propto R^{-5/2}\leqslant 0$ & 0       & $\propto R^{-1/2}$ & Slope                 & Slope                  & $\propto R^{-1}$ \\
 Inner corona & $\propto R^{-3/2}$ & $\propto R^{-5/2}$ &                     $0.1$ & $0$       & $\propto R^{-1/2}$ & Flux                     & Flux & 0 \\
 Outer disk   & $\propto R^{-3/2}$ & $\propto R^{-5/2}$ & Outflow$\leqslant0$          & Outflow & Outflow            & $\nabla\cdot\vec{B}=0$ & $\propto R^{-1}$       & $\propto R^{-1}$ \\
 Outer corona & $\propto R^{-3/2}$ & $\propto R^{-5/2}$ & Outflow$\geqslant0$          & Outflow & Outflow            & $\nabla\cdot\vec{B}=0$ & $\propto R^{-1}$       & $\propto R^{-1}$ \\
 \hline
\end{tabular}
\end{table*}

We prescribe a poloidal magnetic field boundary condition along the boundary, that is defined by satisfying the divergence-free condition.
For that, we need to prescribe the $\theta$-component (only).
At the inner disk boundary we prescribe a fixed inclination of the poloidal magnetic field, with an angle
\begin{equation}
    \varphi = 70\degree\left[1+\exp\left(-\frac{\theta-45\degree}{15\degree}\right)\right]^{-1},
\end{equation}
where $\varphi$ is the angle between the magnetic field and the initial disk surface.

At the inner coronal boundary we require the conservation of the magnetic flux.
The $B_{\phi}$ component is recovered by extrapolating a power law in the radial direction $\propto R^{-1}$.

For the outer boundary conditions, the density, pressure and toroidal magnetic field are computed by imposing a power law extrapolation, 
while the three velocity components follow a zero-gradient prescription.
However, in addition we require that there is no radial inflow into the coronal region and no outflow from the disk region.
Finally, the component $B_\theta$ is computed assuming a power-law $\propto R^{-1}$, while the radial magnetic field is computed 
satisfying the solenoidality condition.

\subsection{The dynamo model}
\label{sec::eqdynamo}%
%
As we have shown recently \citep{2020ApJ...900...60M}, the non-isotropic disk dynamo model of 
\citet{1995A&A...298..934R} proved to have some considerable advantages, e.g. 
the reduced number of parameters required to describe the dynamo tensor and the greater stability 
(compared to a scalar dynamo model) when the initial magnetic field has a non-zero vertical component.
Therefore, in this paper, we apply the dynamo tensor derived by \citet{1993A&A...269..581R,1995A&A...298..934R} 
within the thin disk approximation, thus with negligible non-diagonal components,
\begin{equation}
    \alpha = (\alpha_R,\alpha_\theta,\alpha_\phi) = -[\bar{\alpha}_0\circ \bar{q}_\alpha]c_sF_\alpha(z)
\end{equation}
where the symbol $\circ$ corresponds to the element-wise product of two vectors, and with the adiabatic sound speed $c_s$ at the disk midplane.
The vector $\bar{\alpha}_0$ determines the strength of the dynamo tensor, and $F_\alpha(z)$ describes the vertical profile of the alpha-effect.
Naturally, we also need to assume a sufficiently high disk ionization.
The vector $\bar{q}_\alpha$ is a generic dynamo-quenching function. 
The specific form of the dynamo quenching will be described in Section \ref{Sec::dynquench}.

As pointed by \citet{1995A&A...298..934R}, the effects of the rotation on the turbulence can be described by the Coriolis number,
\begin{equation}
    \Omega^* = 2\Omega\tau_c
\end{equation}
where $\Omega$ is the frequency of revolution and $\tau_c$ is the turbulence correlation time, which can be recovered only by direct 
simulations (see e.g. \citealt{2010MNRAS.405...41G,2015MNRAS.446.2102N,2015ApJ...810...59G,2022arXiv220209272G}).

The exact connection between the local shearing box simulations and the large-scale mean-field dynamo is still unclear, since the values found 
by the local approach ($\Omega^*\lesssim 1$) are $\simeq 10$ smaller than the ones required in order to recover the amplitude of the 
mean-field dynamo.
Future multi-scale dynamo simulations will hopefully solve this problem.

In order to investigate the effect of strong and weak dynamos, we select three values of the Coriolis number.
With $\Omega^* = 10$ we investigate the strong dynamo regime, while $\Omega^* = 5$ refers to the moderate dynamo regime,
and $\Omega^* = 1$ is the weak dynamo regime.
For comparison, all of our simulations are performed in all these three regimes.

The explicit form of $\bar{\alpha}_0$ is in cylindrical coordinates \citep{1995A&A...298..934R,2020ApJ...900...60M}:
\begin{equation}
\label{eq::alphacoeff}
    \begin{array}{lcl}
    \alpha_{0,r} & =  & \DS\frac{1}{2\Omega^{*3}}\left(\Omega^{*2} + 6
    -\DS\frac{6+3\Omega^{*2} - \Omega^{*4}}{\Omega^*}\arctan\Omega^*\right) \\ 
      \noalign{\medskip}
      \alpha_{0,z} & = & \DS\frac{1}{2\Omega^{*3}}\left(-\DS\frac{10\Omega^{*2}+12}{1+\Omega^{*2}}
      + \DS\frac{2\Omega^{*2}+12}{\Omega^*}\arctan\Omega^*\right)\\ \noalign{\medskip}
      \alpha_{0,\phi} & = & \alpha_{0,r}.
    \end{array}
\end{equation}
In the thin disk approximation we can safely assume $\alpha_{0,R} \simeq \alpha_{0,r}$, however, for the sake of consistency, 
all the dynamo vector components are transformed into spherical coordinates.

The profile function $F_\alpha(z)$ confines the dynamo within the accretion disk 
\begin{equation}
    F_\alpha(z) = \left\{
    \begin{array}{lc}
    \sin\left(\pi\DS\frac{z}{H}\right) & z \leq H \\ \noalign{\medskip}
    0 & z > H
    \end{array}
    \right.
\end{equation}
A sinusoidal function (as in \citealt{2001A&A...370..635B}) is adopted instead of a linear function (as in \citealt{1995A&A...298..934R,2000A&A...353..813R})
in order to avoid sharp discontinuities between the disk and the coronal area.

\subsection{The diffusivity model}
\label{sec::eqdiffusivity}
%
The magnetic diffusivity tensor is assumed to have a diagonal structure.
As in \citet{2007A&A...469..811Z,2012ApJ...757...65S}, we adopt an $\alpha$-prescription
\begin{equation}
  \eta = (\eta_R,\eta_\theta,\eta_\phi) = \bar{\eta}_0v_AHF_\eta(z)
\end{equation}
where $v_A$ is the Alfv\'en speed 
at the disk midplane, $H$ denotes the geometrical disk height, and $F_\eta(z)$ describes the vertical profile of the magnetic diffusivity.
As pointed out by \citet{2014ApJ...793...31S}, all  the diffusivity models that we investigate can be represented as
\begin{equation}
    \eta_\phi = \alpha_{\rm ss}c_sHF_\eta(z)
\end{equation}
where $c_s$ is the adiabatic sound speed at the disk midplane.
The diffusivity profile is anisotropic, following
\begin{equation}
    \eta_R = \eta_\theta = \eta_\phi\DS\frac{\eta_{0,R}}{\eta_{0,\phi}}
\end{equation}
\citep{1995A&A...295..807F}
The quantity $\alpha_{\rm ss}$ represents the dimensionless parameter measuring the strength of turbulence \citep{1973A&A....24..337S}.
Implicitly, the magnetic diffusivity is assumed to have a turbulent nature, caused by the MRI \citep{1991ApJ...376..214B}.
We apply
\begin{equation}
    \label{eq::ass_mid}
    \alpha_{\rm ss} = \eta_{0,\phi}\sqrt{\DS\frac{2\mu|_{\pi/2}}{\Gamma\mu_0}}
\end{equation}
where $\mu|_{\pi/2}$ is the magnetization derived along the disk midplane.
However, the magneto-rotational instability can be triggered by both the poloidal and the toroidal field \citep{2013EAS....62...95F}.
As the latter vanishes along the mid-plane by construction, 
 we adopt the following prescription, if not specified otherwise,
\begin{equation}
    \label{eq::ass_d}
    \alpha_{\rm ss} = \eta_{0,\phi}\sqrt{\DS\frac{2\mu_D}{\Gamma\mu_0}}
\end{equation}
where the magnetization $\mu_D$ is defined by the average total magnetic pressure over the geometrical disk height and by the mid-plane
gas pressure.
This approach has also been adopted by \citet{2014ApJ...796...29S}, although a different diffusivity model was applied.

For the strength and the anisotropy of the diffusivity tensor we follow 
\citet{1994AN....315..157K,1995A&A...298..934R,2000A&A...353..813R,2020ApJ...900...60M},
\begin{equation}
\begin{array}{lcl}
      \eta_{0,R} & = & \DS\frac{3}{4\Omega^{*2}}\left[1+\left(\DS\frac{\Omega^{*2}-1}{\Omega^*}\right)\arctan\Omega^*\right] \\ \noalign{\medskip}
      \eta_{0,\theta} & = & \eta_{0,R} \\ \noalign{\medskip}
      \eta_{0,\phi} & = & \DS\frac{3}{2\Omega^{*2}}\left[-1+\left(\DS\frac{\Omega^{*2}+1}{\Omega^*}\right)\arctan\Omega^*\right]
    \end{array}
    \label{eq:eta-tens}
\end{equation}
This is the same model as in \citet{1994AN....315..157K,2020ApJ...900...60M}, although the terms are arranged differently\footnote{following \citealt{2020ApJ...900...60M}} for the sake of clarity. 

Quite different models for the anisotropic diffusivity have been employed in the last decades 
(see e.g. \citealt{2000A&A...353.1115C,2013MNRAS.428..307F}).
In our approach, the strength of anisotropy is not an independent quantity, but depends directly on the Coriolis number.
In the low dynamo regime, $\Omega^*\to0$, the diffusivity is quasi-isotropic.
The anisotropy becomes larger when the Coriolis number reaches higher values.

We apply a profile function
\begin{equation}
    F_\eta(z) = 
    \left\{\begin{array}{ll}
      1 & z\leq H \\ \noalign{\medskip}
      \exp\left[-2\left(\DS\frac{z-H}{H}\right)^2\right] & z > H
    \end{array}\right.
\end{equation}
that allows a smooth decrease of the diffusivity outside the disk region.

\subsection{Dynamo quenching models}
\label{Sec::dynquench}
%
The dynamo action can be understood as a macroscopic effect of the local magneto-rotational instability, which results in an 
additional hyperbolic term in the induction equation.
As the presence of a strong large-scale magnetic field is able to suppress the MRI, the same will happen 
to the dynamo process in the accretion disk as soon as the dynamo-amplified magnetic field becomes strong enough.
Dynamo action will then be quenched.

\subsubsection{Diffusive dynamo quenching}
The Diffusive Dynamo Quenching (DDQ) has been proposed by \citet{2014ApJ...793...31S} in the context of jet launching simulations from 
resistive accretion disk and by \citet{2014ApJ...796...29S} in order to saturate the dynamo amplification of the magnetic field.
With this approach, no direct quenching on the dynamo or the magnetic diffusivity is applied.

Instead, the infinite exponential increase of the magnetic field is prevented by a strong increase in the magnetic diffusivity.
In contrast to the {"}standard{"} diffusivity models (e.g. \citealt{2002ApJ...581..988C,2007A&A...469..811Z}) applied in non-ideal simulations of jet launching regions, 
the quantity $\alpha_{\rm ss}$ is defined as follows,
\begin{equation}
    \alpha_{\rm ss} = \eta_{0_\phi}\sqrt\frac{2}{\Gamma}\left(\frac{\mu_D}{\mu_0}\right)^2.
\end{equation}
The main advantage with this choice of quenching mode is that it avoids the accretion instability \citep{2009MNRAS.392..271C}, 
which may suppress the jet launching process and, in addition, is prone to numerical issues.
On the other hand, the strong dependence of $\alpha_{\rm ss}$ on the magnetization may lead to un-physically high 
values of the magnetic diffusivity.

\subsubsection{Standard dynamo quenching}
So far there is no general consensus about how to calculate the critical magnetization value for the quenching.
Since the turbulence that is causing  the turbulent dynamo effect is supposed to be a consequence of the MRI, the saturation of the
dynamo action should most probably depend on the relative magnetic field strength at the disk mid-plane.

Moreover, a quenching based on the disk magnetization (see e.g. \citealt{2021ApJ...911...85V})
is in agreement with the fact that the MRI is excited by both the poloidal and the toroidal magnetic field\footnote{Note that in case
of hemispheric symmetry, the $B_\phi$ vanishes along the midplane by definition.}.
Thus, we start with the most simple approach for an isotropic quenching model (henceforth Standard Dynamo Quenching, SDQ) \citep{1977SvA....21..479I,2005PhR...417....1B,2016A&A...588A..18M} 
that is basically depending depending on the disk magnetization,
\begin{equation}
    q_\alpha = \DS\frac{1}{1 + \mu_D/\mu_0}
\end{equation}

We point out that such a global - thus non-local - quenching prescription is not easy to fully parallelize in the code.
As investigated by \citet{2021ApJ...911...85V}, a weak parallelization in the $\theta-$direction (i.e. a parallelization 
with only a few number of cores in the $\theta-$direction) overcomes this problem with only little additional computational 
costs.
For a local quenching (using the local magnetization value on the grid cell, see \citealt{2014ApJ...796...29S,2020MNRAS.491.2346T}), 
that is straightforward to parallelize, the main idea of turbulence generation in the disk mid-plane in combination
with the generation of a large scale magnetic flux gets somehow lost. 
Also, the dynamo process itself becomes unstable when every grid cell applies a different strength of the dynamo - again a conflict
with the aim of generating a large-scale magnetic flux.

\begin{figure}
\centering
\includegraphics[width=0.49\textwidth]{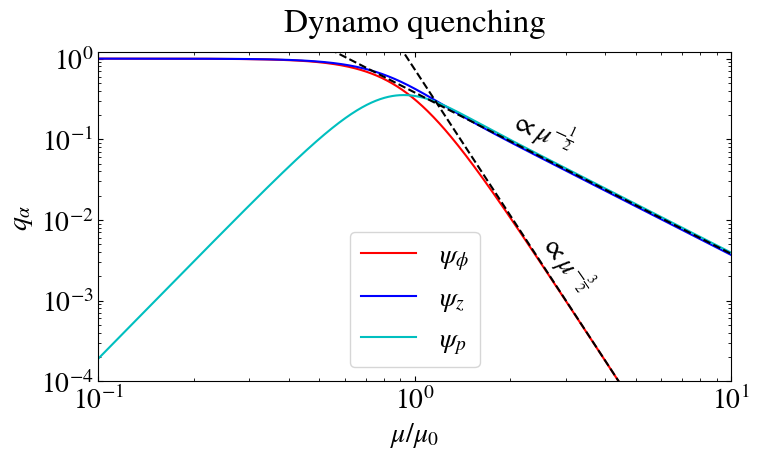}
\caption{The quenching model of \citet{1993A&A...269..581R} for the the different dynamo components as a function of the disk magnetization}
\label{fig::quench_dynamo}
\end{figure}

\subsubsection{Non-isotropic dynamo quenching}
\label{sec::non_iso_dyn}
As for the strength of the mean field dynamo, also the feedback of the magnetic field on the dynamo cannot always be approximated 
with a single scalar function.
Thus, similar to the definition of an an-isotropic dynamo tensor, the quenching of the dynamo effect is also tensorial,
thus acting in different strength on the dynamo tensorial components.

Here, we consider a feedback model (henceforth Non-isotropic Dynamo Quenching, NDQ) that follows from an analytical study 
of turbulence \citep{1993A&A...269..581R}, 
which has elaborated different quenching functions for different components of the $\alpha-$tensor.
It does not only depend on the strength but also on the orientation of the dynamo-amplified magnetic field.

For our purpose, the suppression of the mean-field dynamo is first computed in cylindrical coordinates \citep{1993A&A...269..581R}
and then converted to spherical coordinates,
\begin{equation}
\begin{array}{lcl}
    q_{\alpha_r} & = & \psi_\phi + \DS\frac{15}{8}\DS\frac{B_{z,\rm{D}}^2}{B^2_{\rm D}}\psi_p \\ \noalign{\medskip}
    q_{\alpha_z} & = & \psi_z - \DS\frac{15}{16}\DS\frac{B_{z,\rm{D}}^2}{B^2_{\rm D}}\psi_p \\ \noalign{\medskip}
    q_{\alpha_\phi} & = & \psi_\phi
\end{array}
\end{equation}
where we have defined
\begin{equation}
\begin{array}{lcl}
    \psi_p & = & \DS\frac{1}{4\beta^4}\left[\beta^2 - 5 + \DS\frac{2\beta^2}{3(1 + \beta^2)} \right. \\ \noalign{\medskip}
           & + & \left.\DS\frac{4\beta^4(3\beta^2-1)}{3(1+\beta^2)^3} + \DS\frac{5+\beta^4}{\beta}\arctan\beta\right], \\ \noalign{\medskip}
    \psi_z & = & \DS\frac{15}{64\beta^4}\left[\beta^2 - 3 + \DS\frac{8\beta^4}{3(1+\beta^2)^2} \right. \\ \noalign{\medskip}
           & + & \left.\DS\frac{3+\beta^4}{\beta}\arctan\beta\right], \\ \noalign{\medskip}
    \psi_\phi & = & \DS\frac{15}{32\beta^4}\left[1 - \DS\frac{4\beta^2}{3(1 + \beta^2)^2} - \DS\frac{1 - \beta^2}{\beta}\arctan\beta\right].
    \end{array}
\end{equation}
The quenching parameter $\beta$ is defined as $\beta = \sqrt{\mu_D/\mu_0}$.
The dependence of the quenching components on the magnetization is shown in Fig.~\ref{fig::quench_dynamo}.

Essentially, the quenching depends on the disk magnetization as $\propto\mu^{-3/2}$ for the radial and toroidal component, 
while follows $\propto\mu^{-1/2}$ for the $\theta$-component.
Therefore, we can expect a more sudden and rapid saturation of the magnetic field at lower magnetization.
We point out that $0 \leq q_{\alpha_z} \leq 1$, regardless of the values of $\mu_D$ or $B_{z,D}$.

\subsection{Dynamo number and turbulence parameter}
\label{Sec::dynnum}
%
In order to investigate a set of different simulations,
the comparison of dimensionless parameters which do not have a strict dependence on the initial parameter space,
plays a key role for understanding the physical evolution.
In the context of the mean-field dynamo simulations the dynamo number ${\cal D}$ plays the essential role when it
comes to understand the efficiency of the dynamo process,

\begin{equation}
    {\cal D} =\DS\frac{\alpha_\phi\Omega H^3}{\eta_{\phi,\rm{D}}^2}.
\end{equation}

The dynamo number is the product of the azimuthal 
Reynolds number
\begin{equation}
    {\cal R}_\Omega=R|\DS\frac{d\Omega}{dR}|\DS\frac{H^2}{\eta_{\phi,\rm D}}\approx |\Omega|\DS\frac{H^2}{\eta_{\phi,\rm D}},
\end{equation}
which defines the balance between magnetic diffusion and $\Omega$-effect, and the magnetic Reynolds number
\begin{equation}
    {\cal R}_\alpha = \alpha_\phi H/\eta_{\phi,\rm D},
\end{equation}
which defines the balance between magnetic diffusion and $\alpha-$effect.
Both processes contribute to the amplification of the magnetic field.
While the component $\alpha_\phi$ is related to the amplification of the poloidal components, 
the $\Omega$-effect amplifies the toroidal magnetic field component.

Note, however, that our approach considers as well an $\alpha$-effect for the toroidal field component,
thus considering formally an $\alpha^2\Omega$ dynamo.
Through the different tensorial components both the toroidal and poloidal components of the magnetic 
field are amplified.
However, in our case, the $\alpha$-effect for the toroidal field is always less efficient than 
the $\Omega$-effect,
due to the strong shear that is present in a Keplerian accretion disk (in comparison to e.g. a star).
We therefore can neglect this effect for our further discussion, and consider only the $\alpha_\phi$
for the magnetic Reynolds number, and, thus, also for the Dynamo number.

High dynamo numbers imply the possibility of an efficient dynamo process, low numbers vice versa.
The critical dynamo number, separating the two regimes, depends on the details of the model 
setup (see e.g. \citealt{1988ApJ...331..416S,1990ApJ...362..318S,1994A&A...283..677T}) and
has to be found by applying a series of parameter runs.
For example, \citet{2005PhR...417....1B} find a critical dynamo number ${\cal D}\lesssim10$, below which the magnetic field 
cannot be amplified.
By connecting galactic dynamo simulations with accretion disk simulations, \citet{2021A&A...652A..38B} find ${\cal D}\lesssim7$.

The size of the system is denoted by $H$, here represented by the disk height.
By construction, the dynamo components vanish at $z = H$. 
We therefore compute the quantity $\alpha_\phi$ at $z = H/2$.

Another key parameter in disk simulations as well as in jet launching simulations is the turbulence parameter $\alpha_{\rm ss}$, 
which parametrizes the strength of disk turbulence, i.e. the disk turbulent viscosity \citep{1973A&A....24..337S}. 
On one hand, it represents a direct link between the disk magnetization and the disk diffusivity (see Eq. \ref{eq::ass_d}), on the other hand it can be recovered both from observations and direct simulations.
As pointed by \citet{2007MNRAS.376.1740K}, observational evidences show that a range $0.1<\alpha_{\rm ss}<0.4$ is required to provide a good description of the behaviour of fully ionized, thin accretion disks.
Nevertheless, numerical simulations of direct turbulence recover values which are an order of magnitude below the observational value.

\section{Dynamo feedback models}
\label{Sec::dyn_quench}
Quenching the dynamo tensor prevents an infinite field amplification.
Different quenching methods lead to a different saturation of the magnetic field.
As mentioned above, quenching of the turbulent dynamo is a physical consequence of the process that produces turbulence.
Ideally, quenching models are derived from first principles of turbulent plasmas.

Very general correlations have been found between the accretion disk magnetization and the jet speed or the jet collimation
\citep{2006ApJ...651..272F,2006MNRAS.365.1131P,2016ApJ...825...14S},
demonstrating that a high disk magnetization is tightly correlated with a high jet velocity.

These correlations can be extended, then linking the strength of the dynamo with the jet speed, as a stronger dynamo 
implies a stronger field amplification \citep{2018ApJ...855..130F,2020ApJ...900...59M,2020ApJ...900...60M}.
For this reason, we do investigate the interplay between the amplified magnetic field and 
the dynamo, and how it affects the jet launching process.

\begin{figure*}
\centering
\includegraphics[width=0.7\textwidth]{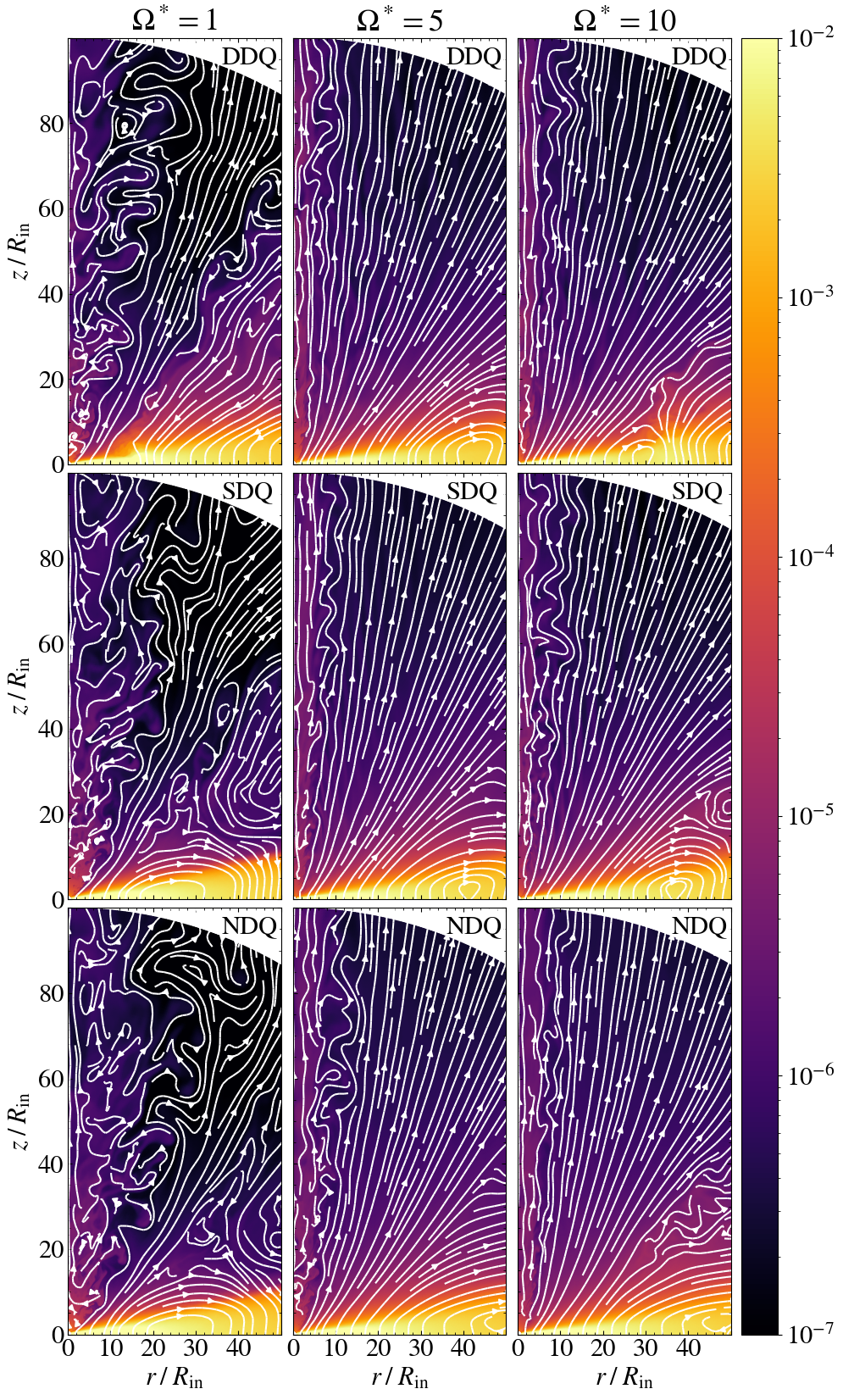}
\caption{Density and magnetic field lines at $t = 10000$ for different dynamo feedback models and Coriolis numbers $\Omega^*$.}
\label{fig::rho_iso}
\end{figure*}

In Figure \ref{fig::rho_iso} we show the density distribution of the disk-jet structure, together with the magnetic field geometry, 
for different feedback models (from left to right, respectively, the diffusive quenching, the standard quenching and the non-isotropic 
quenching) and different Coriolis numbers (from top to bottom, respectively, $\Omega^* = 1,5,10$).
Overall, we see that the magnetic field, near the rotation axis, the magnetic field amplified by the dynamo, has evolved into a large-scale open geometry.
The magnetic field structure, together with its amplification, leads to a highly collimated outflow.

However, for lower Coriolis numbers we notice a more turbulent outflow from the inner disk region.
Such a magnetic field distribution suggests that the outflow is driven by the toroidal magnetic pressure gradient rather than by magneto-centrifugal forces.
These simulations show a major magnetic loop, whose distance from the inner disk depends on the Coriolis number and on the feedback model.
In addition, another loop may emerge, then indicating the presence of a dynamo inefficient zone \citep{2020ApJ...900...59M}.

\subsection{Magnetic field amplification}
%
The primary effect of the dynamo tensor is the amplification of the magnetic field.
However, below a critical value of the Coriolis number, even in presence of a dynamo effect, the magnetic field is
not amplified or is only weakly amplified.
As a direct consequence, for example a fast and collimated outflow cannot be launched.

We identify a critical value for the Coriolis number 
as the one where the dynamo timescale is longer than diffusion timescale \citep{2018MNRAS.477..127D,2018ApJ...855..130F},
\begin{equation}
    \tau_\alpha = \DS\frac{H}{\alpha} > \DS\frac{H^2}{\eta} = \tau_\eta.
\end{equation}

Note that quenching methods may act in quite a different way, 
as the initial strength of diffusivity for the diffusive quenching is $\sim2-3$ orders of magnitudes weaker
than the one from the standard quenching.
This difference strongly reflects on the existence of a critical Coriolis number.

When applying the diffusive quenching method, we have recovered a critical value (i.e. the value below which the magnetic field is not immediately amplified) of the initial Coriolis number $\Omega^*_C \simeq0.15$ \citep{2020ApJ...900...60M}.
On the  other hand, when applying the standard quenching method, or the non-isotropic quenching methods we have developed,
a critical value of the Coriolis number (in order to determine whether the initial dynamo can amplify the magnetic field)
$\Omega^*_C\simeq2$ is found.

\begin{figure}
\centering
\includegraphics[width=0.49\textwidth]{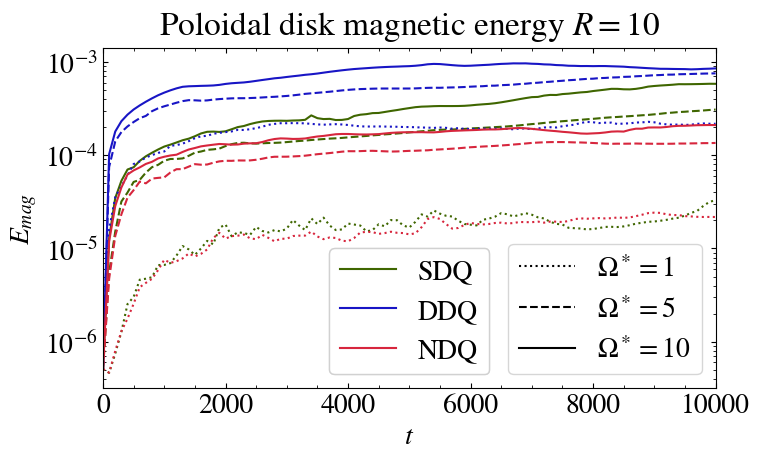}
\caption{Evolution of the poloidal  magnetic field disk energy from $R=10$ to $R = 100$ for different feedback models and Coriolis numbers.}
\label{fig::bpol_iso}
\end{figure}

This difference can be seen in Fig.~\ref{fig::bpol_iso}, where we show the time evolution of the poloidal magnetic energy from radius $R = 10$ to the end of the domain ($R_{\rm{out}} = 100$), while applying a Coriolis number of $\Omega^* = 1$ (dotted lines).
As expected, the poloidal magnetic energy of the diffusive quenching method, 
which is shown in Fig.~\ref{fig::bpol_iso}, is amplified stronger and more rapidly than for the cases of the other quenching models.

Moreover, the field amplification is preceded by a short decrease.
The reason behind this is that, at $t=0$, the diffusive timescale is shorter than the dynamo timescale.
Thus, the magnetic field is diffused away, leading to a decrease in the magnetization and, therefore, of the magnetic diffusivity.
Once the magnetic diffusivity has decreased, the dynamo timescale becomes again shorter than the diffusive time scale.

When the Coriolis number is higher than its critical value  (i.e. in all the cases considered but the ones defined by the dotted red and green lines in Fig.~\ref{fig::bpol_iso}), the amplification of the magnetic field occurs almost instantly.
Thus, when the magnetic field increases, also the dynamo quenching increases, suppressing the dynamo action and lowering the amplification 
of the magnetic field.
In order to disentangle the impact of the quenching methods we applied Coriolis numbers of $\Omega^* = 5$ and $\Omega^* = 10$.
We find that the field amplification is faster and also stronger when the diffusive quenching model is applied (see Fig.~\ref{fig::bpol_iso}).

We notice that the magnetic energy that originates from dynamo action by employing the diffusive quenching method and a Coriolis number 
$\Omega^* = 1$ is approximately the same of the one obtained by the standard quenching method and $\Omega^*=10$ until $t = 3000$.
Since the diffusive dynamo quenching model does not involve a suppression of the dynamo tensor, the magnetic field saturates as soon as 
the diffusivity is strong enough to counterbalance the dynamo effect.
On the other hand, the standard quenching method features both an increase of the magnetic diffusivity and a decrease of the dynamo.
Thus, the same Coriolis number will lead to a different strength of the disk poloidal field depending on the feedback model.

\begin{figure}
\centering
\includegraphics[width=0.49\textwidth]{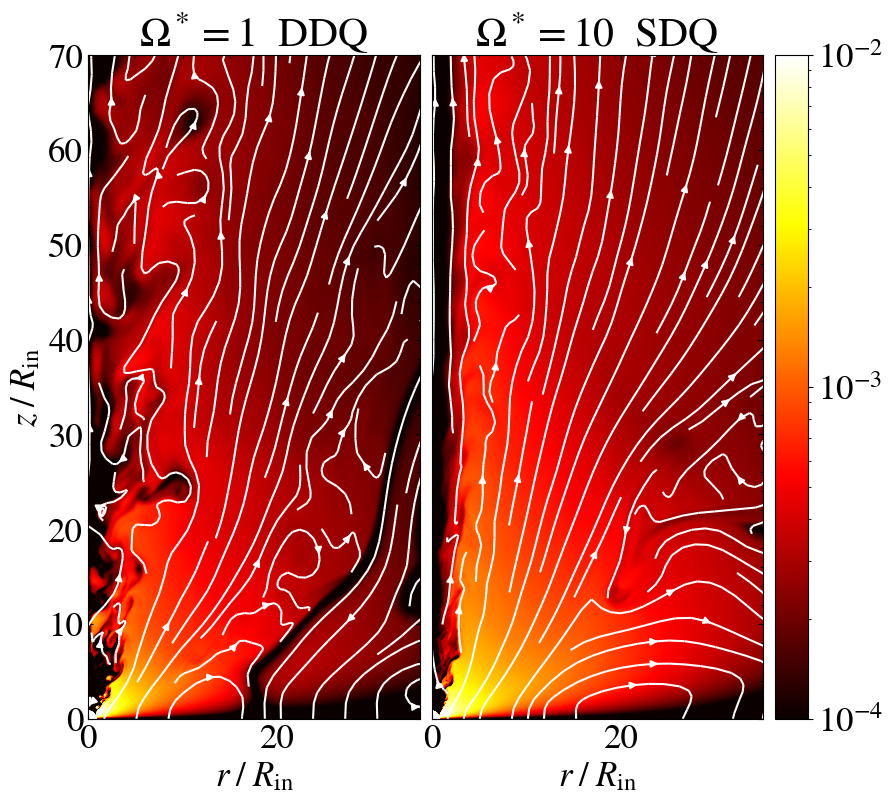}%
\caption{Snapshot of the toroidal magnetic field at $ t = 3000$ for the standard dynamo quenching case and $\Omega^* = 10$ (left panel) and for the diffusive dynamo quenching case and $\Omega^* = 1$ (right panel).}
\label{fig::btor_iso}
\end{figure}

However, as shown in Fig.~\ref{fig::btor_iso}, 
the presence or the absence of the dynamo inefficient zones from the early stages \citep{2020ApJ...900...59M} plays a key role in the jet structure and evolution.
We find that a low Coriolis number leads to the formation of dynamo inefficient zones regardless of the quenching model, in agreement with \citet{2020ApJ...900...60M}.
On the other hand, we also find that dynamo inefficient zones are present for large $\Omega^*$, which we did not find in our previous works.
The formation of these zones may be connected with the increase of grid resolution that is coupled with the HLLC Riemann solver we now apply,
which together provides a better resolution of the disk substructures (probably smeared out by the more diffusive HLL solver).

Finally, the results obtained by applying the non-isotropic dynamo quenching model show no difference from the standard dynamo quenching model 
for low Coriolis number.
However, for higher Coriolis numbers, the different suppression of the dynamo in the non-isotropic dynamo quenching model ($\alpha_\phi\propto\mu_D^{-3/2}$, 
while $\alpha_\phi\propto\mu_D^{-1}$ in the standard dynamo quenching model) 
leads to a different saturation of the magnetic field.
More specifically, we find that the magnetic field, in the non-isotropic dynamo quenching model, is saturated towards a lower disk magnetization than the one 
obtained by the standard dynamo quenching model.

\subsection{Dynamo number and turbulence parameter}
%
The dynamo number is traditionally used to indicate the strength and efficiency of dynamo activity.
A high dynamo number indicates an efficient dynamo, thus leading to strong field amplification.
Vice versa, a low dynamo number indicates that dynamo cannot act efficiently anymore, and the magnetic field generated 
has reached its saturation value - either 
established by a strong magnetic diffusivity, thus by diffusive quenching (diffusive dynamo quenching),
or by suppressing the dynamo activity itself, thus by direct quenching of the dynamo-alpha (standard dynamo quenching, 
non-isotropic dynamo quenching).
The critical dynamo number (see Sec. \ref{Sec::dynnum}) differentiates the two regimes.

\begin{figure}
\centering
\includegraphics[width=0.49\textwidth]{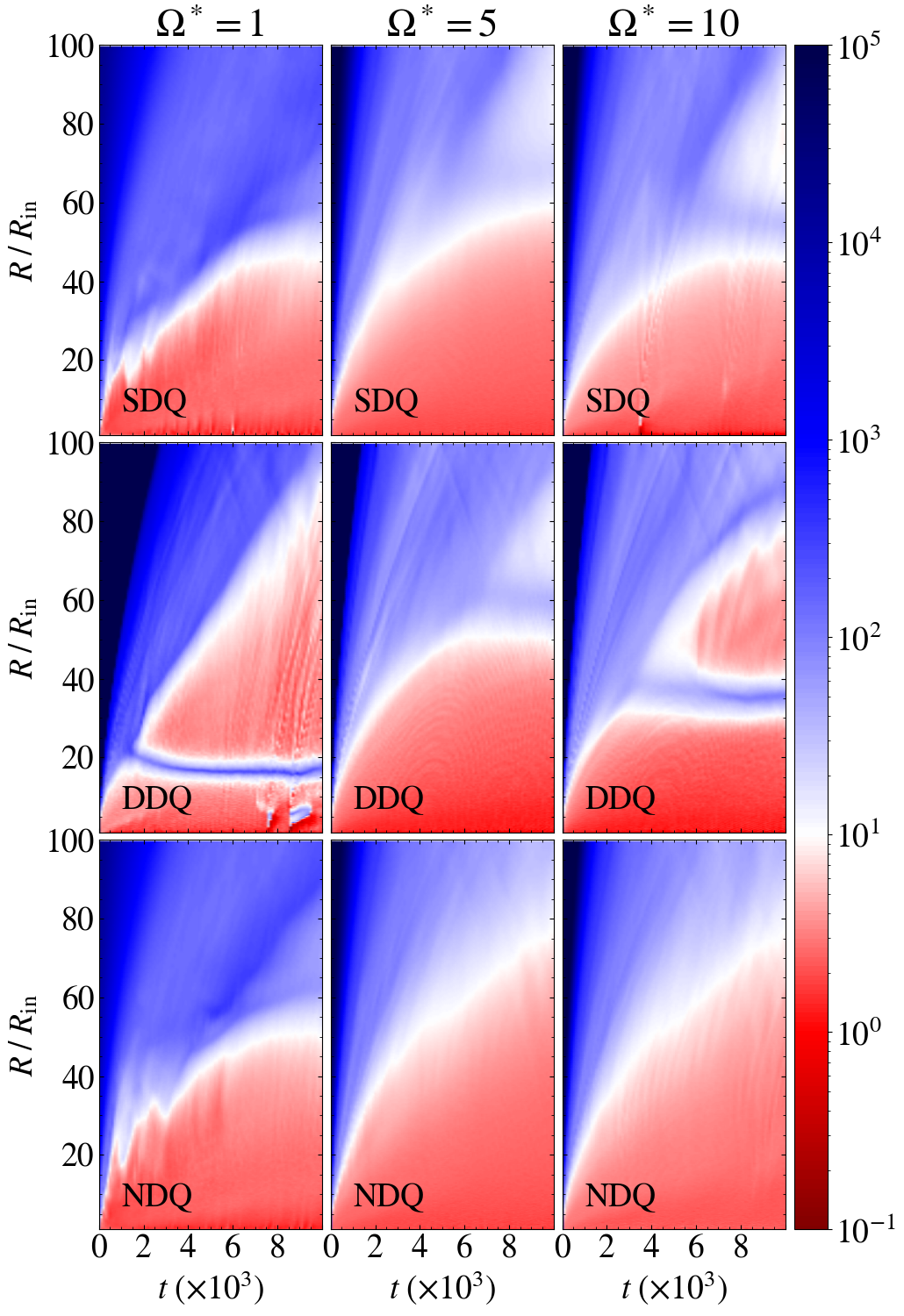}
\caption{Dynamo number $\cal D$ as function of time and radius for different Coriolis numbers and 
feedback models.
}
\label{fig::dynnum_iso}
\end{figure}

Our simulations (see Fig.~\ref{fig::dynnum_iso}) confirm earlier predictions of \citet{2005PhR...417....1B,2021A&A...652A..38B}, that is that
(i) field amplification does not occur for ${\cal D}\lesssim 10$ (red in Fig.~\ref{fig::dynnum_iso}), 
while (ii) amplification occurs when the dynamo number supersedes its critical value since the magnetic diffusivity decreases (blue in Fig.~\ref{fig::dynnum_iso}).

Moreover, we find that the critical value of the dynamo number does not depend on the feedback model applied or the Coriolis number that is given.
This suggests, essentially, that the amplification and the saturation of the magnetic field is a very general property of the mean-field 
dynamo approach, and does not depend on certain modeling details.
Note, that the exact value of the critical dynamo number can be influenced by the numerical resolution applied and 
the numerical algorithms adopted 
(see, e.g. \citealt{1988ApJ...331..416S,1990ApJ...362..318S,1994A&A...283..677T}).

The main difference between the quenching models we apply, is the absence of dynamo inefficient zones for low values of the Coriolis 
number and the feedback models which imply a suppression of the dynamo.
A possible explanation is that a magnetic field reversal can be maintained only if the dynamo does not vanish. 
If the magnetic field is not constantly amplified by the mean-field dynamo (because of the quenching), the field reversal zones are able 
to reconnect and be diffused away.
This is not possible if the dynamo is not suppressed.
However, since the standard dynamo quenching model and the non-isotropic dynamo quenching model lead to a suppression of the dynamo
(and not in the diffusivity), the dynamo inefficient zones are more likely to be suppressed.
We also point out that the presence of dynamo-inefficient zones, which are not strictly related to the component $\alpha_\phi$, is 
still possible.
In this regard, the dynamo number may require a different definition 
considering all the tensorial components of the dynamo.

\begin{figure}
\centering
\includegraphics[width=0.49\textwidth]{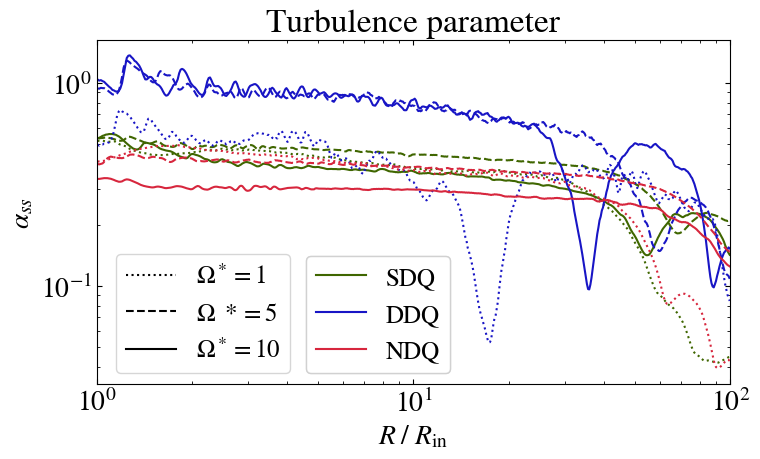}
\caption{Turbulence parameter $\alpha_{\rm ss}$ for different values of the Coriolis number and different feedback models at time $t = 10000$ as function of radius.}
\label{fig::ass_iso}
\end{figure}

On the other hand, the turbulence parameter $\alpha_{ss}$ \citep{1973A&A....24..337S} does not explicitly depend on the dynamo tensor, but only on 
the magnetic diffusivity.
Therefore it can be applied as a useful tool in order to understand the evolution of the magnetic diffusivity once the magnetic field 
amplification took place.
As we can see from Fig.~\ref{fig::ass_iso}, the feedback models which include the suppression of the dynamo and the 
diffusive dynamo quenching model show several differences with regard of how the turbulence parameter depends on the Coriolis 
number once the magnetic field is saturated.
In particular,  we find that the results applying the diffusive dynamo quenching model show a  unique dependence on the Coriolis number.

As pointed out in the previous section, for the diffusive dynamo quenching model, the magnetic diffusion is the only process that is  
able to saturate the mean-field dynamo.
Because of the different amplification of the magnetic field for different Coriolis numbers, 
different in both the strength of the magnetic field and the timescale of amplification,
the disk magnetization saturates to different levels (see Fig.~\ref{fig::bpol_iso}).
Because of the strong, i.e. quadratic, dependence of the diffusivity $\eta$ on the disk magnetization $\mu$, 
in the diffusive dynamo quenching model this dependence is reflected in the fact that also the $\alpha_{ss}$ is found to depend on $\mu$.

\begin{figure*}
\centering
\includegraphics[width=0.95\textwidth]{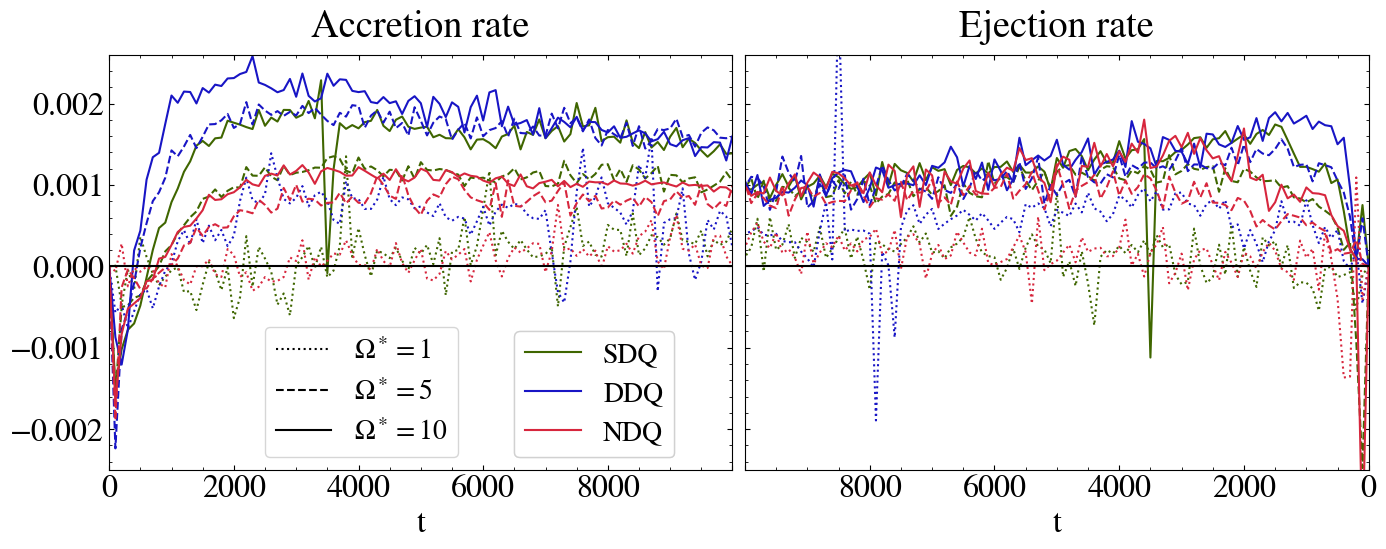}
\caption{Accretion $\dot{M}_{\rm{acc}}(R)$ and ejection rate $\dot{M}_{\rm{eje}}(R; \theta_S)$ for $R = 7$. Note that the $x-$axis of the ejection rate plot is flipped in order to directly compare accretion and ejection at time $t = 10000$}
\label{fig::accrejec_iso}
\end{figure*}

\subsection{Accretion and ejection}
%
The different saturation levels for the magnetic field strength play a key role in the  dynamical evolution of 
the accretion disk and the subsequent jet launching process.
This is demonstrated in Fig.~\ref{fig::accrejec_iso} showing the accretion and the ejection rate of the launching area.
The accretion rate is computed by integrating the net radial mass flux through the disk at $R = 7$,
\begin{equation}
    \dot{M}_{\rm{acc}}(R) = 2\pi R \DS\int_{\pi/2}^{\pi/2 - \theta_S}  \rho v_R R d\theta,
\end{equation}
with the disk opening angle defined as $\theta_S \equiv \arctan(2H/r)$.
Similarly, the ejection rate is computed by integrating between $R_{\rm in}$ and $R = 7$ along the disk surface,
\begin{equation}
  \dot{M}_{\rm{eje}}(R; \theta_S) =  \DS\int_{R_{\rm{in}}}^{R} \rho v_{\theta}(\tilde{R}) 2\pi \tilde{R} d\tilde{R}.
\end{equation}
The interrelation between the accretion rate and the Coriolis number confirms, regardless of the feedback mode, 
previous results obtained by \citet{2018ApJ...855..130F}.
In particular, we find that a stronger dynamo leads to a stronger accretion rate.
We investigated the impact of the magnetic field topology on the accretion process
(i.e. the presence or the absence of dynamo inefficient zones) in detail previously \citep{2020ApJ...900...59M}.

This influence is also confirmed by comparing the standard dynamo quenching model for $\Omega^*= 10$ with the diffusive dynamo quenching model for $\Omega^* = 1$.
Here, despite a similar amplification of the poloidal magnetic field (see Fig.~\ref{fig::bpol_iso}), the presence (DDQ model, $\Omega^* = 1$) or absence  (SDQ model, $\Omega^* = 10$)
of the dynamo inefficient zones, plays a key role in the accretion of material.

On the other hand, the mass ejection, acting on much shorter timescales, shows less pronounced differences for the variety of quenching
methods or the different values of the Coriolis number.
However, we observe that lower $\Omega^*$ generally lead to a higher ejection-to-accretion ratio,
which is in good agreement with the findings of \citet{2020ApJ...900...60M}.
In case of a strong dynamo, the isotropic quenching models show that $<50\%$ of the accreted material is ejected.
This is in good agreement with previous resistive jet launching simulations that do not consider dynamo action, but start
from a prescribed large-scale magnetic field (see e.g. \citealt{2007A&A...469..811Z, 2012ApJ...757...65S}).
For lower Coriolis numbers, accretion requires more time to be established because of the slower amplification of the magnetic field.
On the other hand, the non-isotropic quenching model shows that almost all the matter accreted is ejected.
This is a direct consequence of the feedback model which leads to a saturation of the magnetic field at lower magnetizations (se Fig.~\ref{fig::bpol_iso}).
As a consequence, the magnetic diffusivity, which depends on the disk magnetization, is less amplified than in the isotropic feedback models, leading to a lower accretion.
For this reason, during certain periods of time, the ejection-accretion rate (defined as $\dot{M}_{\rm ej}/\dot{M}_{\rm accr}$)
may actually exceed unity.
This may imply that disk volume regions of very low mass or density may be present for some time until they are replenished from the 
mass reservoir at larger disk radii.

We observe a particularly interesting period at time $\approx3500$ for the standard quenching model (STQ) with high 
Coriolis number $\Omega^* = 10$.
A sudden drop in the accretion and, consequently, in the ejection rate appears.
When looking at the strength of the mean-field dynamo $\phi-$component as a function of radius, 
we notice that at $t = 3500$ it becomes significantly stronger than immediately before or after.

\begin{figure}
\centering
\includegraphics[width=0.49\textwidth]{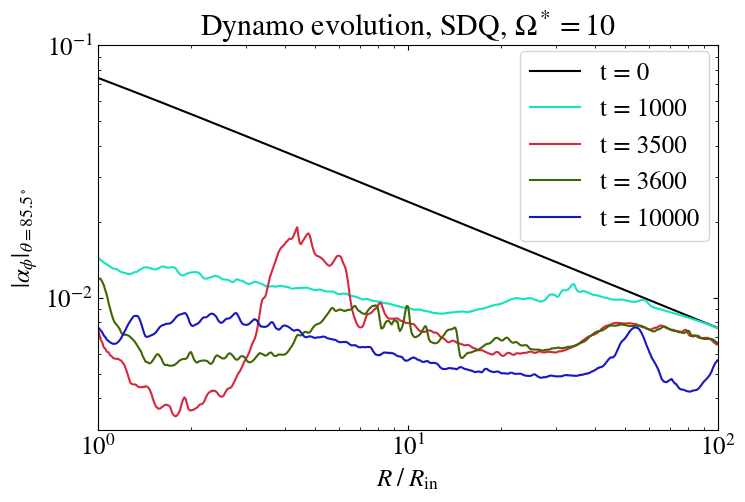}
\caption{Evolution of the dynamo at each radius for the $\Omega^* = 10$ and standard dynamo quenching model simulation.
The angle $\theta = 85.5^\circ$ corresponds to the thermal disk height (and to half of the geometrical disk height).
At this angle the angular profile of the dynamo reaches its maximum absolute value.
}
\label{fig::dynamotime_iso}
\end{figure}

We find that the reason behind this sudden change is a small decrease in the toroidal magnetic field strength,
which seems to amplify the dynamo because of the smaller quenching term (which is related to the magnetic field strength, see Fig.~\ref{fig::dynamotime_iso}). 
In particular, a weaker magnetic field (and therefore a weaker magnetization) leads to a lower quenching and therefore a stronger amplification of the magnetic field.
In addition, low magnetic field strength triggers a lower magnetic diffusivity, which, in turn, curbs the accretion process, because it implies a lower diffusivity.
Once the magnetic field is amplified again by the dynamo, the system goes back to a more stable configuration.

\subsection{Jet properties}
\label{sec::jet}
%
All our models, which consider a feedback of the magnetic field on the dynamo, lead to a quasi-steady, saturated state.
Therefore, our dynamo simulations should retrieve, qualitatively, the essential jet properties found by simulations that do not invoke a dynamo process 
(see e.g. \citealt{2009MNRAS.400..820T,2010A&A...512A..82M,2016ApJ...825...14S}), 
regardless of the method for the dynamo feedback.

In \citet{2020ApJ...900...60M} we discovered a unique numerical correlation between the Coriolis number $\Omega^*$ 
(and therefore the dynamo strength) and the asymptotic jet speed.
In this paper, we now  make a further step in this regard and investigate this interrelation for
different feedback models.
For this purpose, we select certain magnetic flux surfaces (thus contours of the vector potential, respectively poloidal magnetic
field lines), and compute the local disk magnetization and the poloidal velocity of the corresponding outflow along that field line.

\begin{figure*}
\centering
\includegraphics[width=0.95\textwidth]{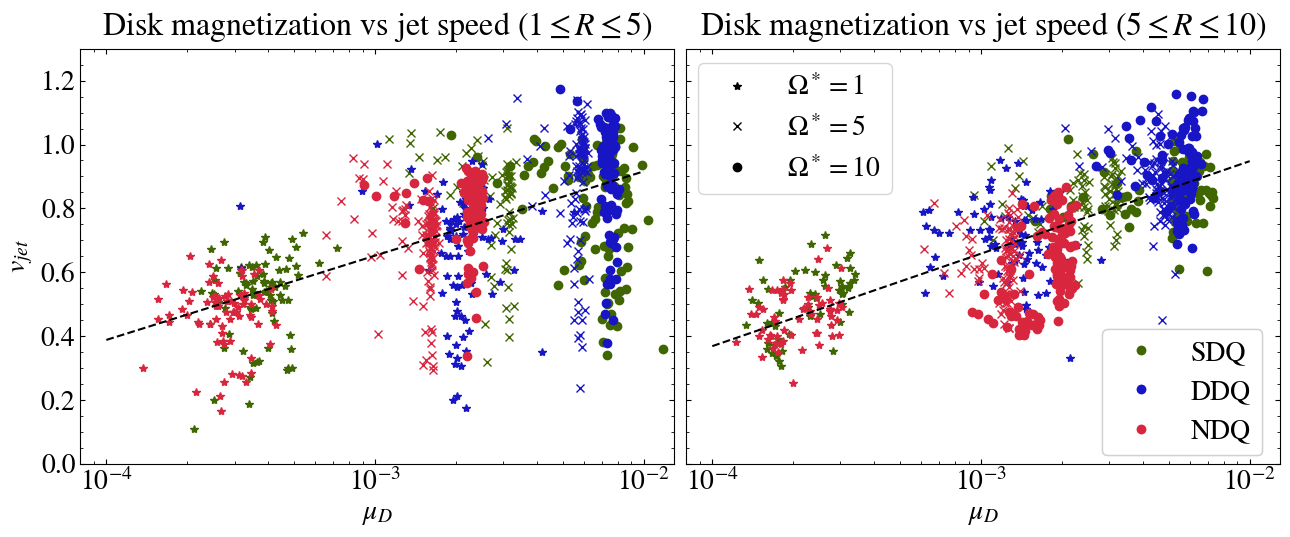}%
\caption{Relation between disk magnetization and jet speed for different feedback models and different values of the Coriolis number. The dashed line represent the extrapolated relation for the three feedback models.}
\label{fig::dmagvel_iso}
\end{figure*}

We do this for each of the code outputs (uniformly distributed and separated by $\Delta t = 100$), starting from $t = 700$, 
i.e. the time when a jetted outflow 
is already formed and has reached the outer boundary, until the final time step of each simulation.
The results are shown in Fig.~\ref{fig::dmagvel_iso}, where the two panels present, respectively, the values obtained for 
radii $1.5 < R < 5$ and $5 < R < 10$.

As we can see, the radial distance makes a difference concerning the smoothness of the interrelation.
For radii $5 < R < 10$ the interrelation looks very well defined , while for smaller radii this correlation is partially broken.
At small radii we notice the presence of vertical {'}columns{'}, i.e. zones with similar magnetization that exhibit a variety
in jet speed.
This is mainly due to the time evolution of these parameters: both the disk magnetization as well as the jet speed
may vary in time for the same radial distance, during the same simulation.
Such variations occur more frequently in the inner disk region because of different timescales involved.

Our understanding is the following.
Small temporal variations of the dynamo leads to suppression and amplification of the magnetic field in the inner disk region (as seen on the left region of Fig. \ref{fig::dynamotime_iso}).
As a consequence, small differences in the magnetic field can lead to different outflow speed in the inner launching region.
Such differences are leading to internal shock within the jet, affecting the outflow.
The small variations in the inner disk magnetic field are also able to affect the ratio between the poloidal and the toroidal magnetic field, which can be related to the jet speed and collimation.
This is a result of the magneto-centrifugal acceleration involved.

On the other hand, at larger radii the system has reached saturation towards a steady state, leading to a 
more narrow interrelation.
At even larger radii, $R > 10$, the low magnetization and the slow rotation lead to a weak disk magnetization and, 
therefore, to a slower outflow speed.
We point out that at these large radii, the disk has accomplished only few revolutions, and the whole
inflow-outflow structure has not yet settled into a quasi-steady state.

However, as a key result, despite showing differences in both the disk magnetization and the jet speed, the different feedback 
models we have examined show in general a unique trend.
They all follow a very similar relation between the two quantities, suggesting that this relation jet speed versus disk magnetization 
does not depend on the dynamo process, the diffusivity model, or the quenching method.
It simply confirms the general relation between these leading inflow-outflow parameters that have been discovered
previously \citep{2016ApJ...825...14S, 2020ApJ...900...60M}.

Still, the feedback of the magnetic field on the dynamo action plays a key role for the saturated disk magnetization.
This holds in particular, because of the more efficient suppression of the dynamo, the non-isotropic dynamo quenching model reaches the saturation 
of the magnetic field already at a lower magnetization levels, about one order of magnitude below.

\section{A consistent quenching mode for diffusivity}
\label{Sec::eta_quench}
%
In the last section we have considered the effects of applying different dynamo feedback modes, 
meaning how to realize the physical effect of quenching the dynamo activity by a strong magnetic field.
Here, we want to go one step further towards a self-consistent modeling of mean-field dynamos.
That is to consider the back-reaction of the magnetic field on the magnetic diffusivity.
Because of their common origin - the turbulence of the disk material -  both the
quenching of the mean-field dynamo and the magnetic diffusivity should be treated in a similar way.
A strong global magnetic field suppresses the turbulence and, thus, both the turbulent dynamo
effect and and the turbulent magnetic diffusivity.

In this section we will put this on more physical grounds, considering a quenching model that follows from analytical 
mean-field theory and that incorporates effects on both the mean-field dynamo and the magnetic diffusivity, 
following the prescriptions of \citet{1994AN....315..157K,1994GApFD..78..247R} (henceforth Consistent Turbulence Quenching, CTQ).

\begin{figure}
\centering
\includegraphics[width=0.49\textwidth]{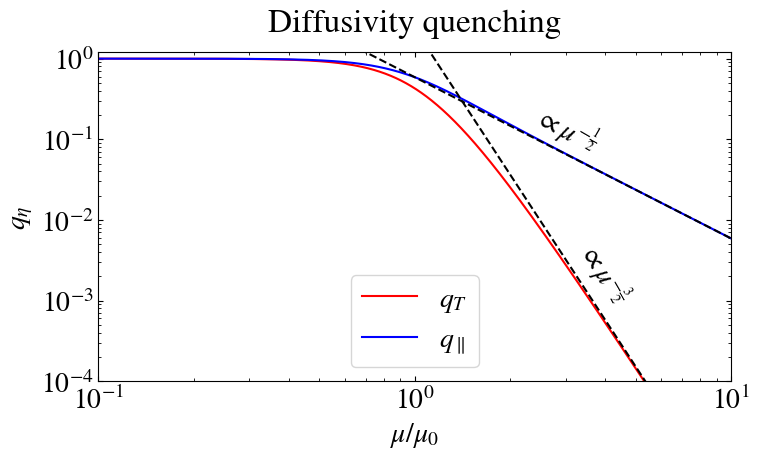}
\caption{The quenching model of \citet{1994AN....315..157K,1994GApFD..78..247R} the $\eta-$diffusivity as function of the disk magnetization}
\label{fig::quench_eta}
\end{figure}

\subsection{Quenching of turbulent magnetic diffusivity}
%
For the feedback of the magnetic field on the dynamo - the dynamo quenching - we follow \citet{1993A&A...269..581R},
as described in Section \ref{sec::non_iso_dyn}.
As in Section \ref{sec::eqdiffusivity} the an-isotropic components of the magnetic diffusivity can be computed directly 
in spherical coordinates.

Here we apply the quenching model following \citet{1994AN....315..157K,1994GApFD..78..247R},
\begin{equation}
  \eta = (\eta_Rq_{\eta_R}, \eta_\theta q_{\eta_\theta}, \eta_\phi q_{\eta_\phi}),
\end{equation}
with
\begin{equation}
    \begin{array}{lcl}
     q_{\eta_R} & = & \DS\frac{3}{2\beta^2}\left[-\DS\frac{1}{1+\beta^2} + \DS\frac{1}{\beta}\arctan\beta\right], \\ \noalign{\medskip}
     q_{\eta_\theta} & = & q_{\eta_R}, \\ \noalign{\medskip}
     q_{\eta_\phi} & = & \DS\frac{3}{8\beta^2}\left[\DS\frac{\beta^2 - 1}{\beta^2 + 1} + \DS\frac{\beta^2+1}{\beta}\arctan\beta\right].
    \end{array}
\end{equation}

The dependence of the diffusivity quenching components on the magnetization is shown in Fig.~\ref{fig::quench_eta}.
We point out that the dependence of magnetic diffusivity on the disk magnetization is also determined by the disk turbulence
parameter $\alpha_{\rm ss}$.
Here, we model this applying $\alpha_{\rm ss} \propto \sqrt{\mu_{\rm D}}$ (as in Eq.~\ref{eq::ass_d}).

Mean-field dynamo models applying a diffusivity quenching (as in \citealt{1994GApFD..78..247R}) have been applied in the 
context of galactic dynamos \citep{1994A&A...286...72S,1996A&A...306..740E}, although the quenching model was never coupled with
a non-isotropic diffusivity.
Here, because of the rapid disk rotation and the strong magnetization needed for jet launching, 
we have included both an-isotropic effects.
In the limits of slow rotation and weak magnetization, the diffusivity tensor becomes isotropic. 
When either a rapid disk rotation or a strong magnetization becomes relevant, the isotropy of the diffusivity tensor is broken.
This is the first time, that such modeling with a higher degree of more self-consistency, has been applied in the context 
of jet launching simulations from accretion disks.

\begin{figure*}
\centering
\includegraphics[width=0.95\textwidth]{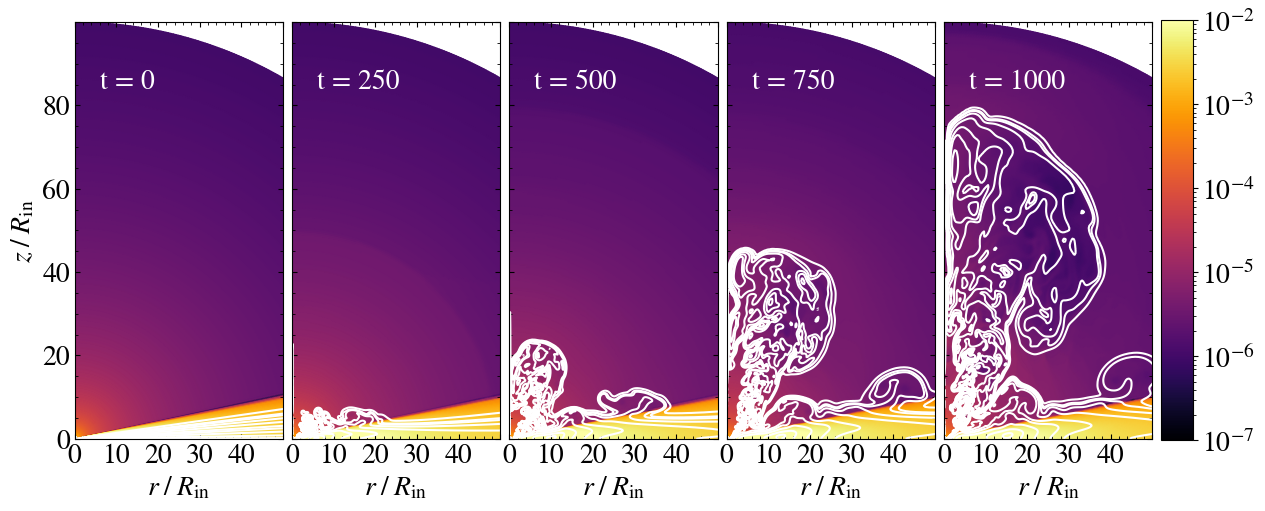}
\caption{Mass density (colors) and magnetic field lines of the reference simulation ($\Omega^* = 1$, CTQ feedback mode) at times $t = 0,250,500,750,1000$.}
\label{fig::rho_an}
\end{figure*}

\subsection{A reference Simulation}
%
In order to investigate the effects of the feedback of the magnetic field on the diffusivity, we have chosen to focus on the 
case $\Omega^* = 1$ as a reference simulation.

In Fig.~\ref{fig::rho_an}, we show the time evolution of the density and poloidal magnetic field of this simulation.
We see that
the saturation of the magnetic field occurs on a short timescale, i.e.  already until $t = 1000$, corresponding
to $\simeq 50$ revolutions of the inner disk.
 At this point in time, the magnetic energy is amplified by an order of magnitude, while the magnetic field lines are
already opened up to a radius $R = 70$ in the outflow region.

Since our approach considers an $\alpha^2$-$\Omega$ dynamo, the toroidal magnetic field is amplified by the disk rotation and by the radial dynamo component \citep{2020ApJ...900...59M}.
Although the dynamo components are quenched by the magnetic field, the $\Omega$-effect still acts on the toroidal magnetic field component.
This, combined with the shorter timescale of the $\Omega$-effect (compared to the amplification of the poloidal field through the dynamo action), leads to a faster and stronger amplification of the toroidal magnetic field.

Note that our initial condition is that of a purely radial field.
However, as in \citet{2018ApJ...855..130F}, we point out that the dynamo-amplified magnetic field should not depend on the initial conditions.

The formation of a magnetic loops (rooted at foot points of different radius in the accretion disk) strongly indicates,
that, at least in the early evolutionary stages, the launching mechanism for this initial outflow is that of a tower-jet,
thus a magnetic pressure driven-outflow (\citealt{1994MNRAS.267..146L,1996MNRAS.279..389L}).
This is further supported by the inclination of the magnetic field towards the disk surface, 
as being not favorable for a magneto-centrifugal driving of the outflow.
As the system evolves, the magnetic loops diffuse outwards, more and more poloidal field lines break up
and the magnetic field geometry reaches the inclination required for a Blandford-Payne-like outflow.
The system evolves further until a quasi-steady state is reached.
At this point, the system consists of a highly magnetized accretion disk and a super-Alfv\'enic 
disk wind, which evolves into a high-speed outflow.
The Alfv\'en surface is located at $\approx5$ thermal disk scale heights above the disk surface.

\begin{figure}
\centering
\includegraphics[width=0.49\textwidth]{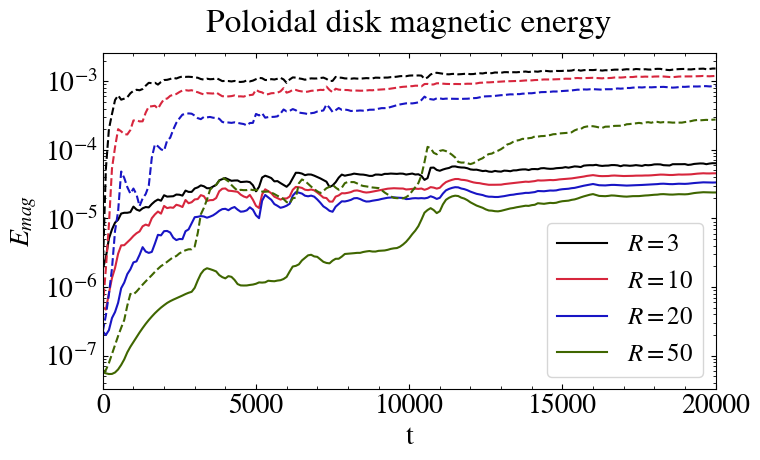}
\caption{Evolution of the poloidal magnetic field disk energy for different disk portions. Solid lines show the poloidal magnetic
energy, while dashed lines show the total magnetic energy (poloidal + toroidal). The radii that are labeled denote the lower integration boundary,
while the upper integration boundary is at the end of the domain, $R = 100$.}
\label{fig::bpolmag}
\end{figure}

\subsubsection{Amplification of the magnetic field}
%
The amplification of the poloidal magnetic energy within the accretion disk is shown in Fig.~\ref{fig::bpolmag}.
The amplification of the magnetic field is stronger in the innermost accretion disk regions, because of the combination of the $\Omega$- and the $\alpha-$effects, in agreement with \citet{2018ApJ...855..130F}

The poloidal magnetic energy is amplified by about 2 orders of magnitude and occurs, mostly, within $t = 3000$.
By comparing the red line of Fig.~\ref{fig::bpolmag} with the dotted green line of Fig.~\ref{fig::bpol_iso}, we notice that the feedback on the diffusivity has a very minor impact on the disk poloidal magnetic field until $t = 4000$. 
However, the oscillations in the magnetic energy at $t \sim 5000$, $t = 6000$, $t = 8000$ and $t \sim 12000$ are a consequence of this novel diffusivity feedback model (see Section \ref{sec::flares}).

 We observe a different evolution compared to our previous models without feedback on diffusivity \citep{2020ApJ...900...60M}.
The magnetic energy does not undergo any intermittent decrease before the magnetic field reaches a quasi-steady state.
We think that this difference can be explained by a combination of effects that rely on the chosen feedback model.
At first, the quenching of the dynamo leads to a lower disk magnetic diffusivity, just
because of the lower magnetization level at which the magnetic field saturates.
Then, because of the lower magnetic diffusivity, the disk mass loss is less compared to a diffusive quenching model 
(where the dynamo tensor is not explicit suppressed), 
therefore the sound speed (which affects both dynamo and the diffusivity) shows no decrease.
Moreover, the magnetic diffusivity is suppressed to an even lower level, just because of the feedback model (quenching
of turbulent magnetic diffusivity). 

At $t \simeq 20000$ the magnetic field has saturated out to a radius $R \lesssim 50$.
The field still continues to be amplified in the outer disk region, as this part of the disk has performed so far
only few rotations and is not yet in dynamic equilibrium.
Also, since the $\alpha-$effect is less efficient here, it takes more time to saturate the dynamo.
Nevertheless, the evolution of the outer disk can weakly affect the disk evolution, 
most likely by triggering episodic ejections (see Section \ref{sec::flares}).

\begin{figure}
\centering
\includegraphics[width=0.49\textwidth]{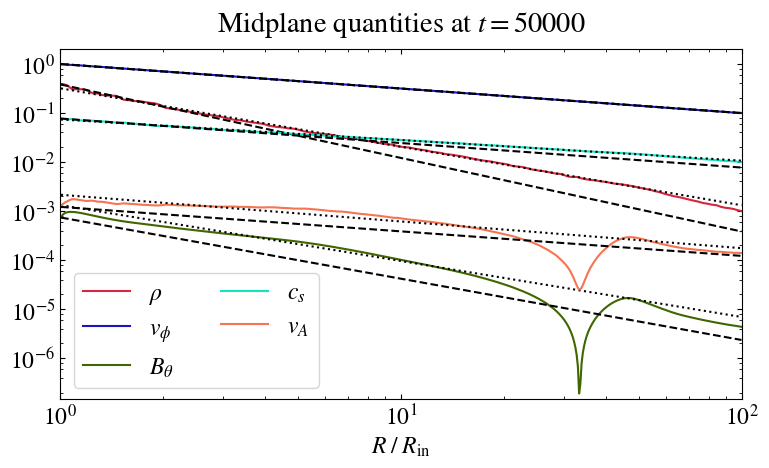}
\caption{Mid-plane quantities at $t = 50000$ for the reference simulation.
The solid lines represent the radial profiles of some selected MHD quantities at the disk mid-plane, 
while the dotted and dashed lines represent, respectively, their corresponding power-law approximation and 
their solution assuming self-similarity.
On the y-axis are the corresponding variables, normalized in code units.}
\label{fig::midplane}
\end{figure}

\subsubsection{Disk structure}
\label{sec::midplane}
%
Because of the magnetic field amplification and the evolution of the magnetic field structure due to dynamo activity, also the disk structure evolves through time.
This is partly due to the change of forces acting on the disk material, but also due to the change of angular momentum balance
and the subsequent re-distribution of disk material.

At time $t = 50000$ the magnetic field has saturated in most parts of the disk (of the size we investigate).
We may therefore investigate the disk structure by looking at the profiles of its mid-plane quantities.
In Figure~\ref{fig::midplane} we display the radial profile of some leading MHD quantities measured at the disk midplane 
and compare them with an idealized radial self-similar solution of the steady-state MHD equations (see \citealt{1982MNRAS.199..883B}).
We also compare with the power law approximation (obtained by performing a linear fit on the logarithmic radial profiles) for each quantity.

We see that the disk kinematics remains unaffected by the dynamo action as the disk rotation remains Keplerian, 
i.e. $\beta_{v_\phi} = -1/2$. 
The density profile power-law index changes, however, from $\beta_\rho = -3/2$ to  $\beta_\rho = -5/4$,
as the mass is mostly accreted from the inner disk, and the whole disk looses only very little mass.

We find that both the sound speed and the Alfv\'en speed show very small deviations from the self-similar solution.
The power-law indexes obtained are, respectively, $\beta_{c_s} \approx -4/9$ and $\beta_{v_A} \approx -5/9$.
The radial dependence of the magnetization that is obviously strongly affected by dynamo action, 
can be recovered by computing the ratio of the Alfv\'en speed vs sound speed,
which corresponds to computing the difference in the respective power-law indexes,
\begin{equation}
    \beta_\mu = \beta_{v_A} - \beta_{c_S} = -\frac{1}{9}
\end{equation}

We find a very good agreement with the results obtained by \citet{2014ApJ...793...31S,2014ApJ...796...29S}, suggesting that the
properties of the saturated state do not depend (or depend  only weakly) on the quenching model for
dynamo action and diffusivity. 

\begin{figure*}
\centering
\includegraphics[width=0.95\textwidth]{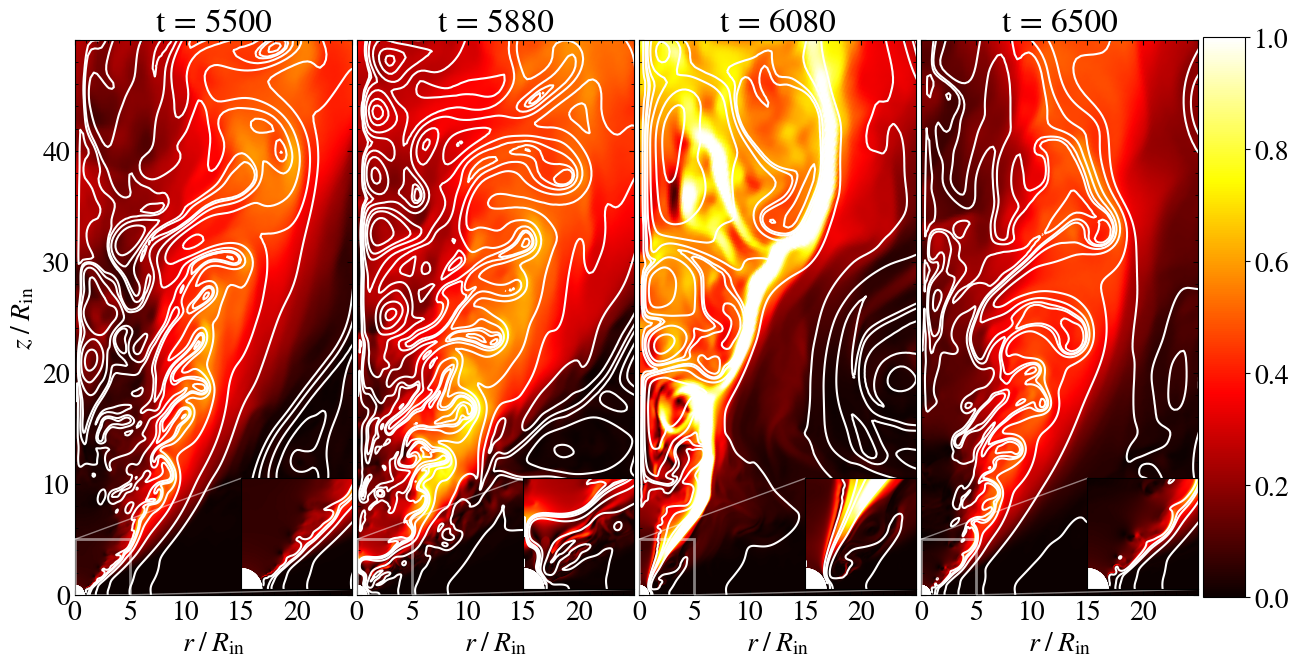}
\caption{Time evolution of the intermittent ejection for the reference simulation ($\Omega^* = 1$, CTQ feedback model). From left to right, snapshot of the poloidal velocity (colors), superimposed with magnetic field lines (defined as the contour lines of the vector potential), are shown at times $t = 5500,5880,6080,6500$.
On the bottom right part of each panel a $5\times5\;R_{\rm in}$ cut from the inner most region is shown, indicating that the jet turbulence is ejected from the innermost disk region.}
\label{fig::flares}
\end{figure*}

\subsubsection{Intermittent ejection}
\label{sec::flares}
%
As pointed before \ref{sec::midplane}, also the system that undergoes consistent feedback concerning the dynamo action and magnetic diffusivity evolves to a quasi-steady state.
However, we diskover an interesting, intermittent feature.
Between $t = 4000$ and $t = 12000$ the magnetic energy features substantial oscillations, especially in the inner disk 
region (see Fig.~\ref{fig::bpolmag}).
We now want investigate the origin of such processes and the consequences on the disk and jet structures.
When the quenching models of Section \ref{Sec::dynquench} are adopted, the feedback applies only to the mean-field dynamo, 
finally leading to a saturation of the magnetic field.
Note, however, that even if the dynamo is quenched, for a low diffusivity it can lead to a re-amplification of the magnetic field.
This is what is happening if the diffusivity follows a consistent quenching model.

We point out that the amplification process of the magnetic field depends not only on the strength of the dynamo 
and the diffusivity tensors, but also on the radial and angular dependence of the tensor components.
This is an essential property of the non-linear quenching models.
The dependence of the magnetic field amplification on the radial and angular profiles for dynamo and diffusivity plays a key role in the dynamo and diffusivity models,
in particular, when the magnetic field shows a reversal within the accretion disk.
Under such circumstances, the magnetic diffusivity is sufficiently high in order to trigger reconnection
processes, but, on the other hand, not strong enough to saturate the magnetic field amplification.
As a result, flux ropes and current sheets are more prone to be formed in the accretion disk,
before they are lifted above the disk surface, similar to what have been described in \citet{2009MNRAS.395.2183Y}.

These flux ropes are able to reconnect above the disk surface in the disk corona, since the given magnetic diffusivity
profile.
Note that also the polarity of the toroidal field is opposite to the one of the launching region of the accretion disk.
As a consequence, these flux ropes, after undergoing magnetic reconnection, are advected towards the accreting
object and are able to disrupt the jet launching.

These features are highly interesting and may have essential relevance for jet launching conditions, such that 
this intermittent behavior may be related to the generation of jet knots.
The understanding of the physical processes behind the flaring activity is not straightforward as they result from the temporal evolution of the highly non-linear resistive MHD equations.

Our understanding is as follows.
The magnetic field geometry emerging from the dynamo activity and the subsequent opening of flux loops enhances the field reversal process typical of the mean-field dynamo action.
While the field structure in these areas are prone to reconnection, the low field strength over there also implies a 
lack of magnetic pressure support.
Further, since the reconnection area is not rooted in the disk via the magnetic field (no lever arm), it
is only slowly rotating.
In combination - lack of pressure support and centrifugal forces - gravity wins and leads to a collapse
of this area towards the central object, thereby cutting off the inner disk wind.

Lateron, because of the opposite polarity, the magnetic field in the inner disk decreases, leading to a restoration of the dynamo.
Then the magnetic field is re-amplified, leading to a strong ejection and acceleration of the disk matter.
As the field is amplified, the quenching on the mean-field dynamo saturates the magnetic field, which goes back to both strength and topology that it had before this episodic fast ejection.

The time period between to consecutive flares increases after each flare, since the reconnection process occurs 
further and further from the launching region.
(the first flare appears at $t = 5000$, the second at $t = 6000$, the third at $t = 7500$ and the last at $t = 10000$).
We expect less reconnecting plasma and thus less variability in the jet once the disk has reached a saturated state.

\subsection{A parameter Study}
%
In order to investigate the similarities and the differences between the consistent turbulence quenching and the dynamo quenching 
methods, we have performed simulation runs applying different Coriolis numbers $\Omega^*$.

\begin{figure*}
\centering
\includegraphics[width=0.95\textwidth]{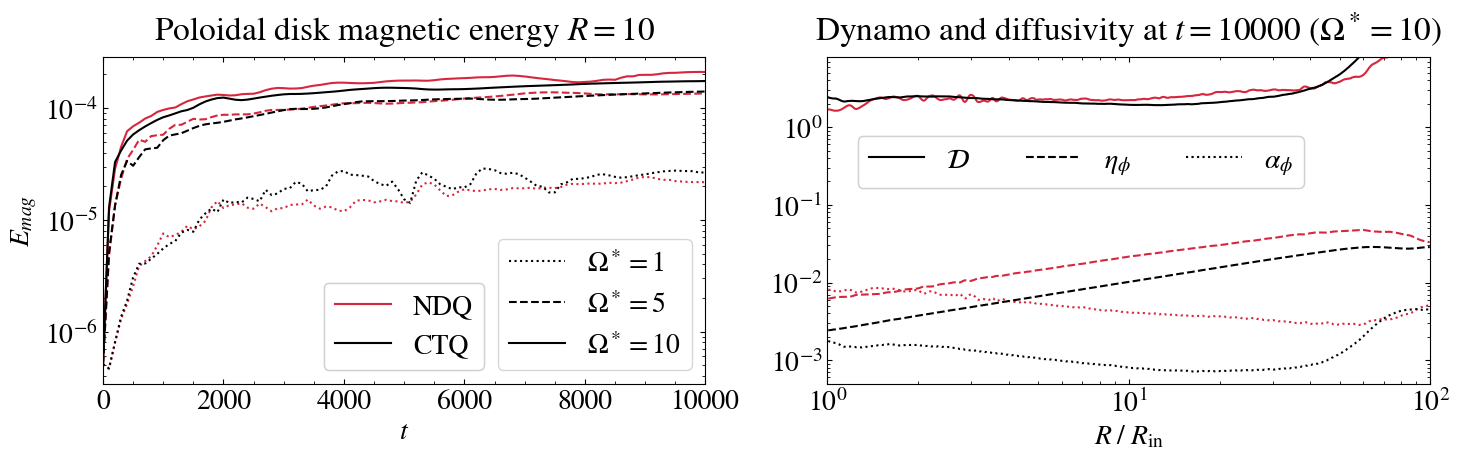}
\caption{Magnetic field evolution for the NDQ and CTQ quenching models. 
In the left panel is shown the 
Time evolution of the disk poloidal magnetic energy.
In the right panel only the cases with $\Omega^* = 10$ are considered.
Shown are the respective radial profiles for dynamo numbers, the $\phi$ components of magnetic diffusivity at midplane, and the mean-field dynamo $\alpha $ at half disk-height , all at $t = 10000$.}
\label{fig::Bpol_comp_an}
\end{figure*}

The results are shown in Fig.~\ref{fig::Bpol_comp_an}.
In the left panel we see the time evolution of the poloidal disk magnetic energy, that is essentially the dynamo-generated
field amplification by the $\alpha$-effect.
We notice that for a higher Coriolis number, the differences between the non-isotropic dynamo quenching model and the consistent turbulence quenching model are only little.
This finding is interesting as it may sound counter-intuitive - wouldn't one expect that a stronger magnetization 
is leading to a stronger quenching on the magnetic diffusivity, and therefore to more differences with the model 
without the consistent quenching?

However, note that a stronger dynamo implies also a stronger feedback on the dynamo (as shown in Section \ref{Sec::dyn_quench}).
Because of the feedback (quenching) on the diffusivity, in combination the $\alpha-$ and the 
$\Omega$-effect lead to a 
stronger amplification of the toroidal magnetic field\footnote{which occurs on a timescale shorter than the 
amplification of the poloidal field}.
Since the dominating launching mechanism at the early evolutionary stages of jet launching is the toroidal pressure-dominated 
launching (tower jet), 
the quenching on the dynamo {\em and} the diffusivity is mainly triggered by the {\em toroidal} field.

As a result, the quenching of the diffusivity plays a minor role in the amplification and saturation of the poloidal field.
Therefore, despite the quenching of magnetic diffusivity, which would be thought to lead to a higher dynamo number,
we find that these numbers are almost identical as shown in Fig.~\ref{fig::Bpol_comp_an} (right panel) for the case $\Omega^*=10$
at $t = 10000$.
This is consistent with the usual understanding that the dynamo number is a useful tool in order to characterize the onset of
field amplification for a dynamo (in the disk).

\begin{figure}
\centering
\includegraphics[width=0.45\textwidth]{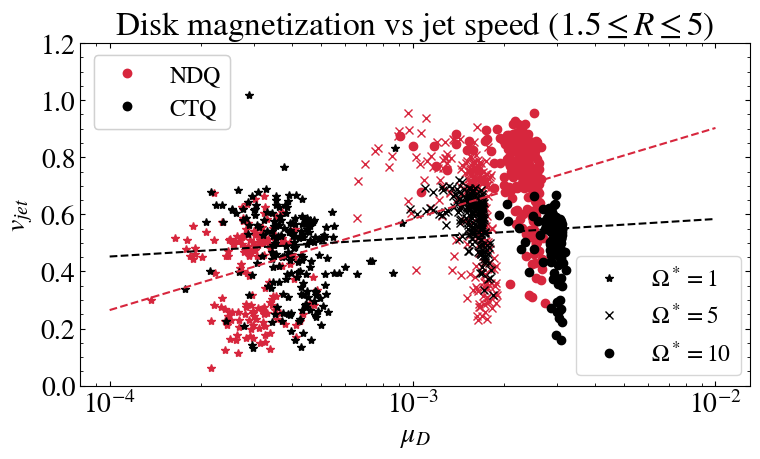}
\caption{Jet speed vs disk magnetization for the NDQ and CTQ model. The dashed lines represent the extrapolated fit for the corresponding feedback model.}
\label{fig::vpolmag_an}
\end{figure}

\begin{figure*}
\centering
\includegraphics[width=0.95\textwidth]{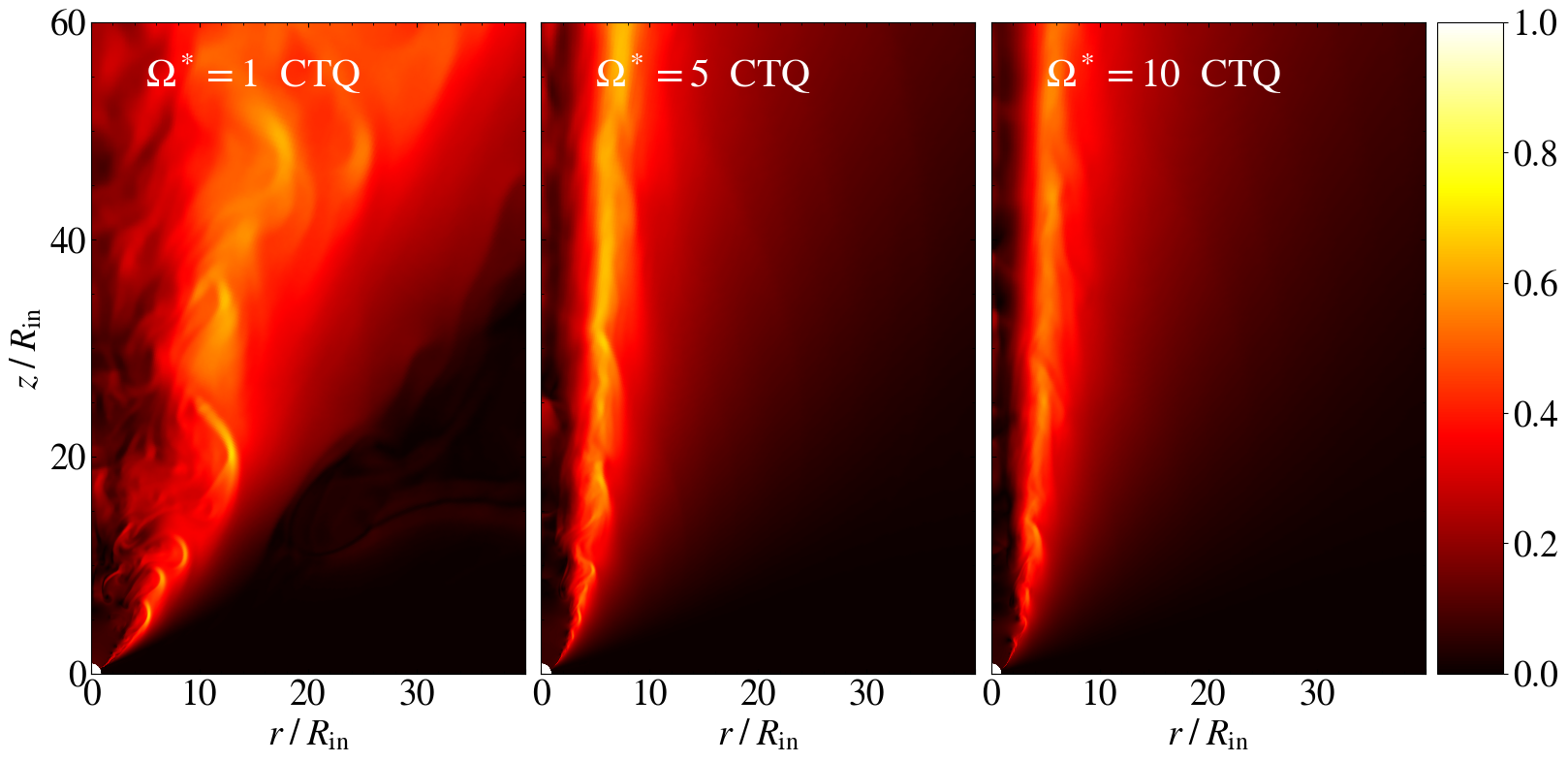}
\caption{Jet speed at $t = 10000$}
\label{fig::vpol_an}
\end{figure*}

On the other hand, although working in the accretion disk, the quenching on the diffusivity plays a key role also for the jet dynamics.
The interrelation between the disk magnetization and the jet speed for the new quenching setup shows substantial differences from those 
by applying only a quenching on the dynamo (see Fig.~\ref{fig::vpolmag_an}).
While we find a clear correlation between the disk magnetization and the jet speed in Section \ref{sec::jet} above, 
the simulations with the consistent feedback thus the quenching of diffusivity show almost no correlation between these two 
quantities.
On the other hand, the simulations with the consistent turbulence quenching and high Coriolis number show an outflow with an extremely
high degree of collimation (see Fig.~\ref{fig::vpol_an}).
This lack of correlation is also present at larger radii. 

We recognize that at $t = 10000$ the magnetic loops, which are the main driver for the magnetic tower at initial evolutionary stages,
are diffused outwards, and the outflow that is launched form the inner disk is now driven by the magneto-centrifugal mechanism,
i.e. launched sub-Alfv\'enic and subsequently supersedes in the Alfv\'en and the fast magnetosonic speed.
However, because we also have strong amplification of the toroidal field, the collimation process is very rapid (in terms of 
spatial scales).
The Alfv\'en surface is rather close to the disk surface where the jet is launched.
As a consequence, magneto-centrifugal acceleration can happen only along a short distance and the final speed
remains rather low.

This is in particular true for the simulations with higher Coriolis number (and therefore stronger amplification of the magnetic field).
The Alfv\'en surface moves further down to the disk surface.
This is a very interesting effect. 
These jets are less efficient concerning the Blandford-Payne acceleration mechanism (when compared with jet launched by dynamo disks with different feedback models, as in, e.g.,  \citealt{2014ApJ...796...29S,2018ApJ...855..130F} or the previous section of this paper), 
but very efficient concerning the Blandford-Payne collimation mechanism.
Note that the latter is indeed a consistent result as the material along collimated field lines cannot be accelerated 
magneto-centrifugally anymore.

In summary, we find that our consistent feedback model that quenches also the disk diffusivity strongly impacts the launching process.
This holds in particular for a weak dynamo (low Coriolis number $\Omega^* = 1$), because of the magnetic field reversals that are induced
in the jet launching region and also the intermittent disruption of the jet followed by the production of a flare.
However, even for high Coriolis numbers, when the differences are less pronounced, the jet structure is severely affected by this new feedback model,
as acceleration decreases and collimation increases.
One may hypothesize about a "critical" or "optimal" Coriolis number for the dynamo that can produce the fastest jets. 
However, this issue need further detailed analysis, which is beyond the scope of this paper.

So far we have investigate only thin Keplerian disks (i.e. neglecting the non-diagonal components of the dynamo tensor).
The differences we find by applying a small Coriolis number (representing a small effect of the rotation on the turbulence) suggest that
we may expect even more structured and probably unstable jets that are produced by {\em thicker} accretion disks.

\section{Conclusions}
\label{Sec:conclusions}
%
In this paper we presented MHD mean-field dynamo simulations in the context of large-scale jet launching.
Our simulations were performed in axisymmetry, evolving all three vector components for the magnetic field 
and velocity.
We employed the resistive MHD code PLUTO 4.3 \citep{2007ApJS..170..228M}, which we have extended by implementing the 
mean-field dynamo action as detailed in \citet{2020ApJ...900...59M}.
 
In extension of our previous works on mean-field dynamo-driven jets \citep{2014ApJ...796...29S,2020ApJ...900...60M}, 
in this paper we have essentially investigated the feedback of the dynamo-generated magnetic field on the 
dynamo activity and the disk diffusivity.
This is a further step towards a consistent numerical modeling of mean-field disk dynamos.

In summary we have applied
(i) different quenching models for the mean-field dynamo, and 
(ii) an analytically derived formalism for the mean-field dynamo and turbulent diffusivity, that consistently incorporates 
the suppression of turbulence by a strong and ordered magnetic field.

The following summarizes our approaches and results.
 
1) We have numerically investigated how different dynamo quenching models affect the magnetic field evolution
and thus the jet launching process.
More specifically, we compared the most common quenching strategies \citep{2005PhR...417....1B,2014ApJ...796...29S} with the 
analytical model of \citet{1993A&A...269..581R}.
The latter model has the advantage to be a more consistent approach, based on an analytical model of turbulent dynamo theory.

2) Essentially, we find that a stronger feedback by the magnetic field on the dynamo leads to a saturation of the magnetic 
field that is generated by the dynamo at lower disk magnetization.
On the other hand, the so-called standard quenching or the diffusive quenching lead to a stronger saturated magnetic field. 
Nevertheless, the launching process and the jet structure that emerges are affected by the possible evolution of the dynamo inefficient zones, and that even for similar values of the disk magnetization.

3) The diffusive quenching model typically leads to a very stable disk-jet connection, just because of the 
continuous production of large-scale magnetic flux. 
However, the strong coupling in the model between the disk magnetization and the magnetic diffusivity, in combination 
with the absence of a quenching term of the dynamo tensor, may potentially lead to unphysically high values of 
diffusivity.
This problem has been solved by applying a more consistent quenching model on the dynamo and a diffusivity model 
where $\alpha_{\rm ss}\propto\sqrt{\mu_D}$.
 
4) In agreement with previous studies, when we apply a feedback only on the dynamo, we recover an interrelation between the disk magnetization and the outflow speed 
regardless of the Coriolis number (i.e. the strength of the dynamo) or the feedback model.
While the feedback model plays a key role for in the saturation of the magnetic field, the relations between the inflow-outflow 
parameters seems to be independent of the dynamo/diffusivity model applied.
This interrelation holds only when no quenching of the magnetic diffusivity is applied (see point 8 of this section).

5) We applied and investigated the effects of a more consistent quenching model which encompasses the suppression of the turbulence 
by a strong ordered magnetic field, for both the dynamo tensor \citep{1993A&A...269..581R} and 
the magnetic diffusivity \citep{1994AN....315..157K,1994GApFD..78..247R}.
Such an approach has never been applied in the context of jet launching by dynamo-generated magnetic fields.

6) We found that, in the early evolutionary stages, the jet is driven by the magnetic pressure.
Once the magnetic field has saturated in the inner disk region, and the magnetic loops are 
opened up and their central part has diffused outwards, 
the magneto-centrifugal Blandford-Payne outflow is produced.

7) We found that, by applying a consistent turbulence quenching model, reconnection processes lead to the formation of flux ropes (with opposite magnetic field polarity respect to the disk magnetic field) that are accreted and disrupt the jet.
As a consequence, the magnetic field is re-amplified in the launching accretion disk region, leading to a very fast intermittent ejection.

8) When applying the quenching model that consistently quenches dynamo and diffusivity, for higher Coriolis numbers, we do not find
the established interrelation between jet speed and disk magnetization.
Instead, the high Coriolis number is associated with a more collimated jet.
The strong toroidal field induced leads to rather short acceleration distances, such that these
jets gain only little speed, but a high degree of collimation.

In this paper we have concentrated on {\em thin} Keplerian disk, an assumption that also constraints the strength of
certain dynamo tensorial components (then strictly related to the non-diagonal components of the mean-field dynamo).
However, we expect that the variety of disks found around astrophysical objects,
the disk thickness may play a key role in the launching process, and may produce a variety in the magnetic field geometry and, subsequently,
in the jets produced.
Our future simulations will investigate the dynamo action on thick disks and will be presented in a forth-coming paper.


\acknowledgements
We thank Andrea Mignone and the PLUTO team for the possibility to use their code.
All the simulations were performed on the ISAAC cluster of the Max Planck Institute for Astronomy.
We acknowledge many helpful comments by the anonymous referee that improved the quality of the paper.

%
%
%
\vspace{2mm}
\bibliographystyle{apj}

\end{document}